\documentclass[12pt,preprint]{aastex}
\shorttitle{OBSERVATIONS OF IGR~J00291+5934}
\shortauthors{TORRES ET AL.}

\begin{document} 

\title{Observations of the 599 Hz Accreting X-ray Pulsar IGR~J00291+5934 during the 2004 Outburst and in Quiescence}
\author{
M. A. P. Torres\altaffilmark{1}, 
P. G. Jonker\altaffilmark{1,2,3},
D. Steeghs\altaffilmark{1}, 
G. H. A. Roelofs\altaffilmark{4},
J. S. Bloom\altaffilmark{5},
J. Casares\altaffilmark{6},
E. E. Falco\altaffilmark{1}, 
M. R. Garcia\altaffilmark{1}, 
T. R. Marsh\altaffilmark{7}, 
M. Mendez\altaffilmark{2,3,8}, 
J. M. Miller\altaffilmark{9,1}, 
G. Nelemans\altaffilmark{4}, 
P. Rodr\'iguez-Gil\altaffilmark{6}}

\altaffiltext{1}{Harvard-Smithsonian Center for Astrophysics, 60 Garden St, Cambridge, MA 02138}
\altaffiltext{2}{SRON, Netherlands Institute for Space Research, Sorbonnelaan 2, 3584 CA, Utrecht, the Netherlands}
\altaffiltext{3}{Astronomical Institute, Utrecht University, P.O.Box 80000, 3508 TA, Utrecht, the Netherlands}
\altaffiltext{4}{Department of Astrophysics, Radboud University, Toernooiveld 1, 6525 ED Nijmegen, the Netherlands}
\altaffiltext{5}{Astronomy Department, University of California, Berkeley, CA 94720}
\altaffiltext{6}{Instituto de Astrof\'isica de Canarias, 38200 La Laguna, Tenerife, Spain}  
\altaffiltext{7}{Department of Physics, University of Warwick, Coventry CV4 7AL}
\altaffiltext{8}{Astronomical Institute, University of Amsterdam, Kruislaan 403, 1098 SJ Amsterdam, The Netherlands}
\altaffiltext{9}{University of Michigan, Department of Astronomy, 500 Church Street, Dennison 814, Ann Arbor, MI 48105, USA}

\begin{abstract} 

We report on optical and near-infrared observations obtained during
and after the 2004 December discovery outburst of the X-ray transient
and accretion-powered millisecond pulsar IGR~J00291+5934. Our
observations monitored the evolution of the brightness and the
spectral properties of IGR~J00291+5934 during the outburst decay
towards quiescence. We also present optical, near-infrared and {\it
Chandra} observations obtained during true quiescence. Photometry of
the field during outburst reveals an optical and near-infrared
counterpart that brightened from $R\simeq 23$ to $R\simeq 17$ and from
$K=19$ to $K\simeq 16$. Spectral analysis of the $RIJHK$ broadband
photometry shows excess in the near-infrared bands that may be due to
synchrotron emission. The H$\alpha$ emission line profile suggests the
orbital inclination is $\simeq 22^{\circ}-32^{\circ}$. The preferred
range for the reddening towards the source is $0.7\leq E(B-V)\leq0.9$,
which is equivalent to $4.06 \times10^{21}$ cm$^{-2}$ $\leq N{_H} \leq
$ $5.22 \times10^{21}$ cm$^{-2}$. The {\it Chandra} observations of
the pulsar in its quiescent state gave an unabsorbed 0.5-10 keV flux
for the best-fitting power-law model to the source spectrum of ($7.0
\pm 0.9$) $\times 10^{-14}$ ergs {cm$^{-2}$} {s$^{-1}$} (adopting a
hydrogen column of $4.6\times 10{^{21}}$ cm$^{-2}$). The fit resulted
in a power-law photon index of $2.4^{+0.5}_{-0.4}$. The $(R-K)_0$
color observed during quiescence supports an irradiated donor star and
accretion disk. We estimate a distance of 2 to 4 kpc towards
IGR~J00291+5934 by using the outburst X-ray light curve and the
estimated critical X-ray luminosity necessary to keep the outer parts
of the accretion disk ionized. Using the quiescent X-ray luminosity
and the spin period, we constrain the magnetic field of the neutron
star to be $< 3 \times {10^8}$~Gauss.

\end{abstract} 


\section{INTRODUCTION}

Most known pulsars are isolated neutron stars characterized by pulse
(spin) periods of about 0.5 s that increase at a rate of
$\dot{P}=dP/dt\sim{10^{-15}}$ s/s as the rotational kinetic energy is
carried away by magnetic dipole radiation (see e.g. Lorimer 2005 and
references therein). This implies young neutron stars with
characteristic ages of $\tau=P/2\dot{P}\sim10^{7}$ years and a
magnetic field strength of $B\propto{(P\dot{P})^{1/2}}\sim10^{12}$
Gauss. However, a small fraction of this pulsar population has
millisecond periods with rotation periods ranging from 1.5 to 30 ms
and ${\dot{P}}\sim10^{-19}$ s/s. They are thought to be old neutron
stars ($\tau \sim10^{9}$ years) with magnetic fields of $B\sim10{^8} -
10{^9}$ Gauss. Binary evolution theory predicts that millisecond
pulsars are spun-up during mass transfer from the companion star onto
the neutron star during the low-mass X-ray binary phase (see
e.g. Alpar et al. 1982; Radhakrishnan \& Srinivasan 1982; Bhattacharya
\& van den Heuvel 1991). This prediction agrees with the wealth of
millisecond pulsars found in binary systems ($\sim 80 \%$ of the
current sample compared to $\lesssim 1\%$ of the young
pulsars). Additionally, eight low-mass X-ray binaries have been found
(to date) to harbor an accretion-driven millisecond X-ray pulsar (see
e.g. Wijnands 2005a). All eight were discovered as X-ray transients
when they underwent an outburst caused by an episode of intense mass
transfer onto the neutron star via an accretion disk.

The accretion-driven millisecond X-ray pulsar IGR~J00291+5934
was first detected in outburst on 2004 December 2
during Galactic plane scans with {\it INTEGRAL} (Eckert et
al. 2004). Reanalysis of the {\it Rossi X-Ray Timing Explorer} ({\it
RXTE}) {\it All-sky Monitor} ({\it ASM}) data archive showed that the
source was likely active during 1998 November and 2001 September,
leading to a tentative recurrence time of approximately 3 years (Remillard
2004). Follow-up observations with the {\it RXTE Proportional Counter
Array} ({\it PCA}) revealed that IGR~J00291+5934 (hereinafter J00291)
has a 147.4 min binary orbit and harbors a neutron star spinning at
599 Hz (1.7 ms). To date this is the fastest neutron star spin
observed for an accretion-powered X-ray pulsar (Markwardt et
al. 2004a,b).

Outburst spectra obtained by {\it INTEGRAL} and {\it RXTE} are
consistent with an absorbed power-law, yielding a photon index of
$\alpha \simeq 1.7-1.8$ and a hydrogen column density of $\geq
2\times10^{21}$ cm$^{-2}$ (Shaw et a. 2005; Galloway et al. 2005). The
{\it INTEGRAL} spectrum was also well fitted by a thermal
Comptonization model with electron temperature of 50 keV and Thomson
optical depth $\simeq 1$ (Falanga et al. 2005; Shaw et al. 2005). In
addition, the {\it RXTE PCA} data could be fitted with a two component
model: an absorbed power-law and a thermal component with $kT \simeq
1$ keV (Paizis et al. 2005), with the thermal component being
interpreted as emission originating on a hot spot on the neutron star
surface. Whereas the power-law photon index showed no evolution during
the outburst decline ($\alpha \simeq 1.7$), the thermal component
decreased to the point that it could not be constrained with the data
acquired 10 days after the outburst onset. A simultaneous {\it RXTE
PCA/Chandra} observation acquired on 2004 December 14 revealed a
spectrum in accordance with a 0.4 keV thermal component (likely
associated with emission from the accretion disk) and a power-law
component with a similar index as above (Paizis et al. 2005). The {\it
PCA} data also revealed that J00291 showed atypical
behavior compared to other neutron star low-mass X-ray binaries
during outburst: the power spectra showed broad-band flat-top noise
with very low break frequencies (0.01-0.1 Hz) as well as the highest
integrated fractional $rms$ variability ($\simeq 50$~\%) found to date
in neutron star systems. These properties are more similar to those
detected in low-hard states of black-hole systems than neutron star
systems (Linares, van der Klis \& Wijnands 2006). J00291 was observed
during quiescence on three occasions with {\it Chandra}. The first
observation was taken a month after its discovery; the second and
third observations were taken 12 and 36 days after the initial {\it
Chandra} observation. The analysis of the X-ray spectrum showed clear
evidence of X-ray variability during quiescence (Jonker et al. 2005).

The optical and near-infrared counterparts to J00291 were identified
with a $R\simeq 17$ and ${K}\simeq 16$ mag variable star (Fox \& Kulkarni
2004; Steeghs et al. 2004) coincident with the reported radio
counterpart (Pooley 2004, Fender 2004). The first spectrum of the
optical counterpart indicated the presence of broad emission lines of
He{\sc ii} and H$\alpha$ (Roelofs et al. 2004; see also Filippenko et
al. 2004). A  binary system with inclination $\lesssim85$ degrees was
supported by the lack of eclipses or dips in the X-ray light curve
(Galloway et al. 2005). The donor star in J00291 is likely a hot
(irradiated) brown dwarf and not a fully degenerate star given the orbital
parameters of the system (Galloway et al. 2005, Falanga et al. 2005)
and the presence of hydrogen emission lines in the optical spectrum
(Roelofs et al. 2004).

The organization of the paper is as follows: we begin by describing in
detail the data acquisition and reduction steps (Section 2). The whole
outburst light curve from the monitoring campaign together with
photometric observations during quiescence is presented in Section
3. In Section 4 we obtain constraints on the photometric variability
of J00291 during outburst and quiescence and in Section 5 we derive a
refined astrometric position for the source. The optical spectrum in
outburst is analyzed in Section 6. In Section 7 we constrain the
reddening towards J00291 and in Section 8 we obtain the optical to
near-infrared spectral energy distribution during outburst.  Section 9
presents a detailed analysis of a new {\it Chandra} observation of
J00291 in quiescence. The results are discussed in section 10 where we
examine key questions such as the orbital parameters and distance
towards J00291. Our conclusions are summarized in Section 11.

\section{OBSERVATIONS AND DATA REDUCTION} 

\subsection{Optical Photometry}

Optical photometry of J00291 was obtained with the following
telescopes (see also Table 1):

\begin{itemize}
\item 
The 1.2\,m telescope at the Fred Lawrence Whipple Observatory (FLWO)
in Arizona. The Minicam mosaic camera was in place and J00291 was
imaged with the FLWO Harris $R$-band filter, which closely
approximates the Johnson-Cousins $R$ broad-band filter. 

\item 
The 0.82\,m IAC80 telescope at the Observatorio del Teide on Tenerife
using the $R$-band Johnson filter and the Thomson 1024 x 1024 pixel
CCD camera.

\item 
The 4.2\,m William Herschel Telescope (WHT) at the
Observatorio del Roque de los Muchachos on La Palma using the Aux Port
imager which carries a $1024\times1024$ TEK CCD. The source was
observed with the Harris $R$ and $I$-band filters with filter transmissions
similar to that of Cousins filters. 

\item 

The 2.5\,m Isaac Newton Telescope (INT) at La Palma using the Wide
Field Camera (WFC), the four thinned EEV 2k$\times$4k CCDs and the
Harris $R$-band filter.

\item 

The 3.5\,m Telescopio Nazionale Galileo (TNG) at La Palma. Cousins $R$
and $I$-band filters were used to acquire images with the DoLoReS
instrument equipped with a $2048\times2048$ Loral CCD.

\item

The MMT 6.5\,m telescope at Mt. Hopkins. Data were obtained with the
Sloan $r'$ filter and the MegaCam CCD mosaic camera (McLeod et
al. 2000).

\item 

The 4.2\,m WHT using the Prime Focus Imaging Camera (PFIP) which
carries two $2148\times4128$ EEV CCDs.  Time-resolved photometry of
J00291 in quiescence was acquired in the (Harris) $R$-band.

\end{itemize}

Integration times ranged from 100~s to 10~min depending on telescope
size, atmospheric conditions and target brightness (see Table 1).  The
images  were corrected  for bias  and flat--fielded in the standard
way using {\sc iraf}\footnote{{\sc iraf} is distributed by the
National Optical Astronomy Observatories.}. We performed PSF-fitting
photometry (Stetson 1987) on J00291 and several nearby comparison
stars.  We also performed a photometric calibration of a set of stars
in the field of view of J00291 using several standard stars from
Landolt plates (Landolt 1992) that were observed with the WHT, the 2.0m
Liverpool telescope on La Palma and the FLWO 1.2\,m telescope. The
magnitudes for J00291 were obtained with differential photometry
relative to these local comparison stars.

\subsection{Optical Spectroscopy}

Ten red and ten blue spectra were acquired on the night of 2004
December 5 using the ISIS double-arm spectrograph attached to the WHT
telescope. J00291 was observed using a 1.5 arcsec wide entrance slit
with the CCDs on each spectrograph arm binned by two in the spatial
direction. The blue arm was used with the R600B grating and the
$4096\times2048$ EEV12 CCD array to yield an useful wavelength coverage of
$\lambda\lambda3500-5200$ with a dispersion of 0.86
\AA~pix$^{-1}$. The red arm was used with the R600R grating and the
$2047\times4611$ MARCONI2 CCD. The useful wavelength range covered the
$\lambda\lambda5400-7100$~interval with a dispersion of 0.88
\AA~pix$^{-1}$. The spectral resolution was about $\simeq 5.5$~pixels
FWHM for both arms.

The raw CCD frames were bias and flat-field corrected with standard
{\sc iraf} routines. The spectra were extracted from each frame with
the {\sc iraf kpnoslit} package. The pixel-to-wavelength calibration
was derived from a cubic spline fit (blue spectrum) and a fifth
order Legendre polynomial fit (red spectrum) to Copper-Argon and
Copper-Neon arc lamp spectra taken at each CCD. The root-mean square
($rms$) deviation of the fit was $< 0.1$~\AA~and $<0.06$~\AA~for
the data acquired with the blue and red arms respectively. No absolute
flux calibration was attempted as the weather conditions were
poor. The individual spectra were re-binned into a uniform velocity
scale of 60.23 km~s${^{-1}}$~pix${^{-1}}$~(blue) and 41.92
km~s${^{-1}}$~pix${^{-1}}$ (red) to be rectified subsequently by
fitting a spline function to the continuum after masking the emission
lines.

\subsection{Near-infrared photometry}

Infrared photometry during the X-ray outburst was obtained from
2004 December 8 to 11 with the 1.3\,m Peters Automated Infrared Imaging
Telescope  (PAIRITEL) at FLWO (Bloom et al. 2006). The camera on PAIRITEL consists of
three $256\times256$ NICMOS3 arrays which image simultaneously a
$8.5'\times8.5'$ field of view in the $J$, $H$ and $K_s$ photometric
bands. The observations consisted of a large number of dithered 7.8\,s
exposures on source. The dithered exposures were first bias and
flat-field corrected to be mosaiced together for each individual visit
(see e.g. Blake et al. 2005).  Total integration times per visit
ranged between 3 to 15 minutes. For each visit, instrumental
magnitudes were extracted  and consequently calibrated relative to the
same nearby 2MASS sources for  all exposures. Photometric error
estimates on the infrared magnitudes are based on a combination of
Poisson statistics and the error contribution of the comparison stars
used for each observation.

J00291 was also observed on two nights with the 3.8m United Kingdom
Infrared Telescope (UKIRT) on Mauna Kea. On December 28 the telescope
was equipped with the UKIRT 1-5 micron Imager Spectrometer (UIST;
Ramsay Howat et al. 2004). $J$, $H$ and $K$-band images were obtained
on the $1024\times1024$ InSb array. Observations consisting of a single
9 point jitter pattern with a 30s exposure at each offset position were
performed using the K98 and J98 broad-band filters, giving a total of
4.5 min on source. The same pattern was used with exposures of 60s
when using the J98 broad-band filter, yielding 9 min on source. The
J98, H98 and K98 filters are the Mauna Kea Observatory NIR Photometric
System filters (see Simons \& Tokunaga 2002 and Tokunaga et al. 2002
for more details). $K$-band observations were made on 2005 January 24
using the UKIRT Fast-Track Imager (UFTI; Roche et al. 2003) equipped
with a $1024\times 1024$ HgCdTe array. The observations were obtained
by jittering the image on 9 different positions. Exposures were 30s at
each position and the jitter pattern was repeated six times yielding
27 minutes of total on-source observing time. The standard star FS103
from the set of UKIRT faint standards was observed for photometric
calibration. During both nights a dark frame was obtained in each
filter at the start of each set of observations. The data were reduced
through the automated {\sc ORAC-DR}
pipeline\footnote{http://www.oracdr.org/} (Allan et al. 2002): the
individual frames from each filter were bad-pixel masked, dark
subtracted, flat-fielded (a flat-field was made using the target
frames) and combined to form mosaics. Absolute calibration of the
$K$-band mosaic obtained with UFTI was performed with the standard star
using a median extinction value of 0.088 mag airmass$^{-1}$ at the
Mauna Kea summit. The resulting absolute magnitudes for the 2MASS
stars in the field differed by $\simeq 0.05 \pm 0.03$~mag. The small
discrepancy is consistent with the expected $2\%$ differences between
magnitudes obtained by calibrating with 2MASS and UFTI standard stars
(Hodking, Irwin \& Hewett 2006; Nikolaev et al. 2000). J00291 was not
detected in any of the UIST mosaic frames. These frames were calibrated
using 2MASS stars in the field and 3-$\sigma$ upper limits to the near-infrared
magnitudes of the source were derived from aperture photometry made to
the field stars.

Details of all the observations are summarized in Table~1.

\subsection{X-ray data: Observations and reduction}

We observed J00291 with the backside-illuminated ACIS-S3 CCD on board
the {\it Chandra} satellite. The observation (ObsID 6570) started on
2005 November 24 09:49:44 UT. The total on-source time was 24671.5
s. We limited the read-out area of the S3-chip to 1/8th of its
original size yielding a smaller exposure time per CCD frame in order
to avoid pile-up. The data were reprocessed in a  standard way using
the ciao 3.3.0 software. We searched the data for periods of enhanced
background radiation, but none were found. Hence, all the data were
used in our analysis. Following Jonker et al. (2005), we
extracted the spectrum of J00291 from a circular region with a 3
pixels radius centred on the best-fit source position as provided by
{\it wavdetect}, whereas the background spectrum was extracted using
an annulus centred on the source position with inner and outer radius
of 10 and 30 pixels, respectively. We detect a total of 143 counts in
the source region and 296 in the background region. This yields 0.3-10 keV
background subtracted source count rates of ($5.6 \pm 0.5$) $\times 10^{-3}$
counts per second. This rate is so low that pile-up is negligible. The
expected background count rate in the source region was $\approx 2$ \% of
the total counts in the source region.

\section{The Outburst Light Curve}

Fig.~1 presents the overall optical/infrared light curve. Each data
point represents the mean magnitude per night at each respective
band-pass. {The discovery optical magnitude ($R\simeq 17.4$; Fox
\& Kulkarni 2004) and the $R$-band photometry reported in Biknaev et
al. (2004) have also been included. Note that the follow-up of J00291
could not be continuous as it was hampered by poor weather conditions
during the winter season in the Northern Hemisphere.

Our first images were taken on 2004 December 8, and showed the source
at $R=18.33\pm0.07$~mag, $I=17.58\pm0.06$~mag, $J=17.1\pm0.1$,
$H=16.7\pm 0.2$ and $K=15.9\pm0.1$. We fitted an exponential to our
first eight $R$-band data points, which describes well the decline in
brightness (the $rms$ from the fit was 0.11 mag). From the fit we
infer a rate of decline of $5.2 \pm 0.4$~d mag$^{-1}$ and thereby an
e-folding time of $5.7\pm0.4$~d in flux. Assuming the transient
started to decay with the same trend after the outburst peak, the
extrapolation of the exponential fit to the time of the discovery by
{\it INTEGRAL} (Eckert et al. 2004, Shaw et al. 2005) yields a peak
optical brightness of $R=17.03\pm0.08$~mag. Our $JHK_s$-band coverage
is not sufficient to determine the flux decay rate at near-infrared
wavelengths.

In Fig. 2 we plot the $R$-band magnitudes during the initial decay
together with the {\it RXTE PCA} outburst light curve. A first account
of the {\it PCA} light curve is given by Galloway et al. (2005) and
Paizis et al. (2005). We digitized their data and plotted them on a
logarithmic scale in Fig. 2. This figure shows that the X-ray light
curve decays exponentially with an initial e-folding time of $8.5 \pm
0.3$~d. An exponential decay is expected to happen in short-orbital
period binaries such as J00291 because the X-ray irradiation is strong
enough to ionize the entire accretion disk (King \& Ritter 1998; see
also discussion). The decay becomes faster after 2004 December 10.2
when the e-folding time becomes $2.6 \pm 0.1$~d. {\it INTEGRAL/ISGRI}
observations (10-100 keV) also show a change in the rate of decline in
the light curve close to 2004 December 10, when the initial 6.6~d
e-folding time decreases to 2.2~d (Falanga et al. 2005). The optical
light curve does not show such a change until 2004 December 15, when
there is indication of a steeper decline in the optical
brightness. The optical data are consistent with a decay with rate $2.2
\pm 0.5$~d mag$^{-1}$ (e-folding time of $2.4 \pm 0.5$~d). 

On 2004 December 19, J00291 was found to be at $R=20.90\pm0.06$~mag and it is
difficult to assess the light curve morphology due to the lack of
coverage after that date. It is possible that the brightness decline
slowed down or/and the source underwent mini-outbursts/re-flares in
the optical before it reached quiescence. In this regard, the
magnitudes reported by Bikmaev et al. (2005; open triangles in Fig. 1)
together with our data seem to be consistent with what could have been
an optical mini-outburst with amplitude $\simeq 0.4$ mag starting about
2005 January 7 and lasting for about a week.  At that time J00291 had
definitely ceased its X-ray activity and settled down into quiescence
at X-ray wavelengths (Jonker et al. 2005). The unreddened $R-I$
colour during the outburst decline was $0.75 \pm 0.09$~mag (2004 December
8), $1.1 \pm 0.2$~mag (2004 December 30) and $0.7 \pm 0.3$~mag (2005 January
7). We cannot draw any conclusion about the evolution of the color
index given the large error bars, which were calculated by adding
quadratically the uncertainties on the $R$ and $I$-band photometry.

By the end of the possible optical mini-outburst and our outburst follow-up
(2005 January 14), the optical magnitude had declined to
$r'=22.7\pm0.1$. Ten days later, J00291 was imaged with UKIRT to derive
a $K=19.0\pm0.1$ magnitude for its near-infrared counterpart in
quiescence. Finally, we observed J00291 with the WHT on 2005 October
25/26 to measure a mean magnitude of $R=23.1\pm0.1$ for the
optical counterpart in quiescence. This yields a total amplitude
for the optical outburst of $\Delta R > 4.8$ mag ($\Delta R = 6.1$ mag from
the above extrapolation) and $\Delta K > 3.1$ mag.

\section{Photometric Variability}

Time-resolved photometry of J00291 over the course of the outburst
decline suggested significant variability on time scales of
tenths of minutes to hours with amplitude $\lesssim 0.3$~mag (Bikmaev et
al. 2005; Reynolds et al. 2005), but gave no indication of the orbital
period. Our photometric data sets were also searched for short-term
photometric variability when more than four data points per night were
available. Quantifying the variability of J00291 requires care, since
some of our data were acquired under variable weather conditions. We
have dealt with this by carrying out photometry of field stars with
similar brightness to J00291 or fainter in order to determine the
significance of our photometry through the standard deviation on their
magnitudes.

Seven images in the $JHKs$ band-passes were made with PAIRITEL on 2004
December 10. The root-mean-square ($rms$) of the J00291 magnitudes is
$0.12-0.15$~mag, not different from the comparison stars in the field of
view. Any $rms$ flux variations of J00291 over the observing
interval cannot be much greater than $\approx 15\%$.

We obtained seven $R$-band images during 2004 December 30 with the
WHT and six $R$-band images during 2005 January 7 with the TNG.  We
constrain the $rms$ flux variation in J00291 to be $\lesssim7\%$
and $\lesssim16\%$ during the small fraction of the orbit covered
during the observations (0.15 and 0.087 orbital cycles for the WHT and
TNG respectively).

Time-resolved photometry of J00291 in quiescence was acquired on 2005
October 25 with the WHT. We obtained five 300s and twenty 600s
$R$-band frames. The $rms$ deviation in the photometry over the 1.41
orbital cycles covered by the 600s images (the data with higher s/n
ratio) is 0.25 mag, two times larger than the $rms$ observed for two
fainter comparison stars. The intrinsic $rms$ flux variability of
J00291 in the $R$-band light curve is constrained to be $\approx 22\%$ by
subtracting in quadrature the $rms$ flux variation of the comparison
stars from the variability observed in J00291. To examine the
variability we used both PSF and optimal aperture techniques to
determine the magnitudes which gave similar values.  We do not see any
evidence for orbital modulation in the photometric variability.

\section{Astrometry} 

The position of J00291 was determined using the higher resolution
images acquired at different epochs during the outburst decline. These
were the images obtained with UKIRT/UFTI (pixel scale of 0.091
$''$/pix), WHT/Aux Port (0.237$''$/pix), TNG/DoLoReS (0.551$''$/pix)
and MMT/MegaCam (0.160$''$/pix). The transformation from pixel to sky
coordinates was computed using the {\sc iraf} tasks {\it ccmap} and
{\it cctran} on 6 to 12 bright stars whose PSFs were not corrupted by
CCD oversaturation effects. The positions for these internal
astrometric reference stars were taken from 2MASS and they have an
accuracy of $0.1''-0.2''$ (Skrutskie et al. 2006). The $rms$ errors of
the astrometric fit were 0.05$''$ and 0.01$''$ (UKIRT), 0.02$''$ and
0.07$''$ (WHT), 0.05$''$ and 0.06$''$ (TNG), and 0.06$''$ and 0.07$''$
(MMT) for right ascension and declination respectively. The error on
the position of J00291 provided by the {\it imcentroid} task was
always $<0.03''$. Using the mean of the four measurements we
determined a refined position for J00291 of
$\alpha$(J2000)=00$^h$29$^m$03$^s$.05 $\pm$ 0$^s$.01 and
$\delta$(J2000)=+59$^o$34$'$18$''$.93 $\pm$ 0$''$.05. The errors
represent the $rms$ of the measurements. The above value is in good
agreement with the positions reported for the optical and X-ray
counterparts (Fox \& Kulkarni 2004; Paizis et al. 2004). It differs
from the position for the radio counterpart (Rupen et al. 2004) by
3.2-$\sigma$ in right ascension, corresponding to an angular offset between both positions of $0.25''$.

\section{The Averaged Spectrum}

As indicated in Fig. 1, our spectroscopic observations took place on
2004 December 5 during the initial decline of the outburst when the
optical brightness was R= $17.72 \pm 0.05$ according to our
exponential fit to the optical light curve (Section 3). The data cover
4.5 contiguous hours of spectroscopy representing 1.8 orbital
cycles. The individual spectra have a signal-to-noise ratio (s/n)
$\lesssim10$ at 4500~\AA~and s/n $\approx 10$ at 6300~\AA. We produced
an average spectrum by assigning optimal weights to the individual
spectra to maximize the s/n of the sum. Fig. 3 presents the weighted
sum. The spectra show the presence of broad Balmer lines up to likely
H$\delta$ in emission. The line profile for H$\alpha$ is
double-peaked. An F-test gives a probability  $> 99.99$\%
confidence that a double-Gaussian fit is better representation of the
H$\alpha$ profile than a single Gaussian fit. Due to the low s/n it is
difficult to assess wheter or not the H$\beta$ and H$\delta$ profiles
are double-peaked as well. The high-excitation He{\sc ii} $\lambda$4686
emission line is present, but we do not detect the Bowen blend (at
$\approx \lambda4640$).  In Table 2 we list the measured line profile
parameters: the velocity shift of each line respect to the rest
wavelength, the centroid of the line and the peak-to-peak
separation (for H$\alpha$ only), the FWHM, the full width zero
intensity (FWZI) and the equivalent width (EW). The values reported
are the mean of the measurements obtained by selecting different
wavelength intervals to set the underlying continuum and the
uncertainties correspond to the standard deviation. Table 2 shows that
the emission line profiles are blue-shifted respect to their rest
wavelength. This shift more likely reflects the systemic radial
velocity (neutron star systems tend to have high systemic velocities;
see e.g. White \& van Paradijs 1996) and/or the presence of a
precessing accretion disk (as observed in short-orbital period X-ray
transients in outburst; see e.g. Torres et al. 2002, 2004 and
references therein). H$\alpha$ is the dominant emission line in
the spectrum with a FWHM of $1340 \pm 10$~km s$^{-1}$ and EW=$6.5 \pm
0.4$~\AA. The velocity separation of the peaks in the averaged profile
is $650 \pm 40$~km s$^{-1}$. He{\sc i} $\lambda$5875 in emission
seems to be detected. In this regard, He{\sc i} $\lambda$6678 was
reported from a single 300s spectrum obtained on 2004 December 12
(Filippenko et al. 2004) when the source brightness was $R=18.93 \pm
0.08$. A visible inspection of this spectrum (see Fig. 1 in Reynolds
et al. 2005) shows that He{\sc i} $\lambda\lambda$5875,7065 emission
lines were also present. We measured the EW and radial velocity of the
individual H$\alpha$ and He{\sc ii}$~\lambda 4686$ emission line
profiles. Neither the EW nor the radial velocities showed significant
modulation with the orbital motion.

The main interstellar features detected are the partially resolved
atomic Na D doublet at $\lambda\lambda 5889.95,5895.92$~(total EW of
$1.1\pm0.1$~\AA) and the Ca{\sc ii} $\lambda\lambda 3933.67,3968.47$
lines (EWs of $0.45\pm0.02$~\AA~and
$0.38\pm0.04$~\AA~respectively). The spectra show also the presence of
diffuse interstellar bands at $\lambda5780$ (EW=$0.4\pm0.1$~\AA),
$\lambda6203$ (EW=$0.22\pm0.02$~\AA) and $\lambda 6284$
(EW=$1.9\pm0.4$~\AA). The broad 6284~\AA~band profile is
contaminated with telluric O$_2$.  Longward of $\approx 6800$ \AA, the
spectra are also contaminated by telluric features.

\section{Reddening towards J00291}

Knowledge of the interstellar extinction is necessary in order to
determine the distance to the source and its spectral energy
distribution. $E(B-V)$ can be estimated in a number of different
ways. From its location in the Galaxy ($l=120.1^{\circ}$,
$b=-3.2^{\circ}$), the expected color excess is $\lesssim 0.80$ mag
according to the average H{\sc i} column ($N{_H}$) obtained by
weighting the $N{_H}$ values within one degree along the line of sight
to the source with the inverse of the distance from the source
position ($N{_H}\lesssim4.66\times10^{21}$ cm$^{-2}$; Dickey \&
Lockman 1990). Here we adopt $N{_H}/E(B-V)=5.8\times10^{21}$ cm$^{-2}$
(Bohlin, Savage \& Drake 1978).  $E(B-V)\lesssim 0.71$ mag using the
all-sky reddening maps based on far-infrared emission at 100~$\mu$m
and 240~$\mu$m from dust (Schlegel, Finkbeiner \& Davis 1998). Note
that both radio and dust maps integrate along the whole line of sight
through the Galaxy and that the reddening maps are expected to have
reduced accuracy for $|b|<6^{\circ}$. Fits to the X-ray spectra
acquired with {\it Chandra} and {\it RXTE}  during the X-ray outburst
(Paizis et al. 2004) provided $N{_H}=4.3\pm 0.4 \times10^{21}$
cm$^{-2}$ (High Energy Transmission Grating Spectrometer spectrum -
HETGS) and $N{_H}=4.3{^{+0.7}_{-0.5}} \times10^{21}$ cm$^{-2}$
(combined {\it Chandra} HETGS and {\it RXTE PCA} spectrum). These
values imply a range of $E(B-V)=0.66-0.86$~mag in extinction. Finally,
a reddening of $E(B-V)=0.8 \pm 0.2$ mag can be derived from the
calibration between reddening and the EW for the $\lambda$5780~diffuse
interstellar band (Herbig 1993). Note that we use only this
interstellar band as it shows a better correlation with reddening than
the other diffuse interstellar bands in our spectrum (see Herbig
1975). Based on the above four independent results, we adopt $0.7\leq
E(B-V)\leq0.9$ ($\equiv 4.06 \times10^{21}$ cm$^{-2}$ $\leq N{_H} \leq
$ $5.22 \times10^{21}$ cm$^{-2}$) as a likely range of the extinction
by giving a lower weight to the result from the $\lambda$5780~DIB,
which is the least precise method for deriving the reddening given the
uncertainties in the E(B-V)/EW relationship and the errors in the EW
measured for this DIB. The lower limit on $E(B-V)$ yields a dereddened
color of ${(R-I)_0}=0.45$ ($R-I=0.75$; Section 3), indicating a
temperature of the order of 5000 K when assuming a thermal origin for
the optical flux. This temperature is lower than the $> 10 000$~K
temperatures expected in the outer parts of the accretion disk during
outburst (see e.g. Hynes 2005) and suggests that $E(B-V)$ is likely
larger than 0.7 mag.

\section{Spectrophotometric Energy Distribution during Outburst}

We acquired simultaneous $JHKs$-band images of J00291 with PAIRITEL
during 2004 December 8 to 11. On December 8, 10 and 11 we also
obtained $R$-band photometry with the 1.2m at FLWO and observed the
source in the $I$-band on December 8. To create spectral energy
distributions (SEDs), the optical and infrared magnitudes were
dereddened using ${A_R}(\lambda{_c}=0.641\mu m)=2.09$ mag,
${A_I}(\lambda{_c}=0.798\mu m)=1.49$ mag, ${A_J}(\lambda{_c}=1.235\mu
m)=0.71$ mag, ${A_H}(\lambda{_c}=1.662\mu m)=0.44$ mag and
${A_K}(\lambda{_c}=2.159\mu m)=0.29$ mag according to the Cardelli,
Clayton \& Mathis (1989) extinction law and assuming a reddening
$E(B-V)$ of 0.8 mag (as discussed in Section 7). Fluxes were
calculated from the magnitudes using the flux calibrations and
effective wavelengths specifications for each filter (Bessel 1990;
Cohen, Wheaton \& Megeath 2003). In Fig. 4 we plot the infrared to
optical SED for three nights. We have also included the near-infrared
data acquired on 2004 December 9 for comparison. The errors show the
uncertainties in the photometry. The $R$-band photometry was not
simultaneous to the near-infrared observations and to account for this
we have overplotted a bar representing the upper limit on the source
variability at this bandpass ($\lesssim 0.16$~mag; Section 4). The
uncertainty in the reddening correction also contributes to the
uncertainty in the SED. The error bars in the bottom of Fig. 4 account
for the effect in the flux at each bandpass due to the $\pm 0.1$ mag
uncertainty in the reddening.

In X-ray transients in outburst the continuum emission will fall
gradually from optical to infrared wavelengths when the thermal
spectrum of the X-ray and/or viscously heated accretion disk dominates
the flux output at these wavelengths (see e.g. Beall et al. 1984,
Vrtilek et al. 1990, Hynes 2005, Russell et al. 2006). Our SEDs show a
break in the expected trend at infrared wavelengths where there is
significant excess flux in the $K$-band during the four nights and
likely  in the $H$-band on December 11 as well. This supports the
presence of another source of near-infrared flux in the spectrum. This
source cannot be emission from the donor star as its contribution to
the near-infrared flux is $< 8 \%$ during outburst. This
estimation is based on the $K$-band magnitudes measured during
outburst and in quiescence (Section 3). The near-infrared excess can
be explained by invoking optically-thin synchrotron emission, which is
expected to contribute to the SED with a component with spectral index
$\alpha < 0$ (see Fender 2006 and references therein). This
non-thermal component has been claimed to explain the excess of
optical/near-infrared flux observed during the outburst of the
accreting millisecond X-ray pulsars SAX J1808.4-3658 ($P_{\mathrm{orb}}$=2 hr,
Wang et al. 2001; Greenhill, Giles \& Coutures 2006), XTE J0929-314
($P_{\mathrm{orb}}$= 43.6 min, Giles et al. 2005) and XTE J1814-338
($P_{\mathrm{orb}}$=4.3 hr, Krauss et al. 2005). The  detection of mid-infrared
optically-thin synchrotron emission from a jet in the neutron star 4U
0614+091 (Migliari et al. 2006) adds support to the
above claim. For illustration purposes, we show in Fig. 4 a power-law
fit ($F{_\nu}\propto{\nu{^\alpha}}$) to the 2004 December 8 data
performed after excluding the flux at the $K$-band. It is clear that
the fit is a poor description of the data: $\chi^2_{\nu}=2.4$ and
furthermore we cannot exclude the very likely presence of flux excess
at $H$, $J$-bands and shorter (optical) wavelengths due to the
non-thermal (jet) component that will make the SED flatter
(redder). The near-infrared unabsorbed flux measured on December 9
($0.30 \pm 0.01$ mJy at $2.159~\mu$m, $0.33 \pm 0.02$~mJy at
$1.662~\mu$m and $0.43 \pm 0.02$~mJy at $1.235~\mu$m) lies above the
radio flux of $0.17 \pm 0.07$ mJy at a frequency of 4.86 GHz measured
that day by Rupen et al. (2004; see also Fender et al. 2004). This
suggests a spectrum with spectral index ($\alpha \geq 0$), implying
a flat or slightly inverted synchrotron optically thick spectrum.

\section{X-ray Data: Analysis}

A month after J00291 went into outburst a 4.7 ks {\it Chandra} ACIS-S
observation detected the source at an unabsorbed flux of $(7.9 \pm
2.5) \times 10^{-14}$ ergs {cm$^{-2}$} {s$^{-1}$} (0.5-10 keV). A
serendipitous 18 ks observation by {\it ROSAT} obtained over 1992 July
26 - August 4 also showed J00291 at a similar flux level, confirming
its return to quiescence within a month. Additional 9 ks and 12.9 ks
{\it Chandra} observations obtained on 2005 January 13 and 2005
February 6 showed the source at an unabsorbed flux of $(7.3 \pm 2.0)
\times 10^{-14}$ ergs {cm$^{-2}$} {s$^{-1}$} and $(1.17 \pm 0.22)
\times 10^{-13}$ ergs {cm$^{-2}$} {s$^{-1}$}, revealing that J00291 is
variable in quiescence (see Jonker et al. 2005 for more details).

In this section, we analyze an additional 24.6 ks {\it Chandra}
observation acquired on 2005 November 24.  We performed a similar
spectral analysis to that presented in Jonker et al. (2005) using {\sc
xspec} (v11.3; Arnaud 1996). We rebinned the source spectrum such that
each bin contains at least 10 counts and used data in the energy range
0.5-10 keV. Due to the low number of source counts per bin, we also
checked the spectral fitting results using the Cash statistic (Cash
1979), the results were consistent with those found using chi--squared
fitting. For all spectral fits the hydrogen column density towards
J00291 was held fixed at $N{_H}=4.6\times 10{^{21}}$ cm$^{-2}$, a
likely value of the hydrogen column density as derived in this paper
(Section 7). Note also that the X-ray light curve shows no
significant variability. A Kolmogorov-Smirnov test gives a probability
of 25 \% for the count rate of the source to be constant.

We began by fitting the spectrum using single component models: a
power-law model, a neutron star atmosphere (NSA) model, and a
black-body model. For the NSA model we fixed the neutron star magnetic
field, its radius and mass at 0 Gauss, 10 km and 1.4 M$_\odot$,
respectively. The temperature and normalization were the only allowed
free parameters. The results of these fits are shown in Table 3. As
can be seen in Table 3, the single-thermal models did not provide an
adequate fit to the data ($\chi^2_{\nu}>3.2$) and yield a temperature
for J00291 considerably higher than the temperature observed for other
neutron star X-ray transients in quiescence. We reject these models on
this basis. A single power-law model was statistically acceptable with
$\chi^2_{\mathrm{red}}$=1.2 for 11 degrees of freedom. In Fig. 5 we
have plotted the spectrum showing the power-law fit. The absorbed
0.5-10 keV source flux from the best-fitting power-law model is ($3.8
\pm 1.0$) $\times 10^{-14}$ ergs {cm$^{-2}$} {s$^{-1}$}, whereas the
unabsorbed flux is ($7.0\pm0.9$) $\times 10^{-14}$ ergs {cm$^{-2}$}
{s$^{-1}$}. We also list in Table 3 the unabsorbed 0.5-10 keV flux
derived from other models fit to the data.  Note here that the X-ray
spectrum of the transient millisecond X-ray pulsars SAX J1808.4-3658
(Campana et al. 2002) and XTE J0929-314 (Wijnands et al. 2005b) are
also consistent with an absorbed power-law spectrum. The source flux
from the best-fitting power-law model is consistent with the values
found using previous {\it Chandra} and {\it ROSAT} observations of
J00291 during quiescence (Jonker et al. 2005)\footnote{Note that
Jonker et al. (2005) used $N{_H}=2.8\times 10{^{21}}$ cm$^{-2}$ in
their work. This value of $N{_H}$ was derived from preliminary
analysis of a 18 ks {\it Chandra} observation acquired at the end of
the X-ray outburst (Nowak et al. 2004) and it has been extensively
used in the literature. In the present work, the power-law index and
temperature model parameters obtained using $N{_H}=2.8\times
10{^{21}}$ cm$^{-2}$ are consistent within the errors with those
obtained using $N{_H}=4.6\times 10{^{21}}$ cm$^{-2}$, whereas the
unabsorbed flux is smaller. For comparison, the absorbed and
unabsorbed fluxes are ($4.3 \pm 0.9$) $\times 10^{-14}$ ergs
{cm$^{-2}$} {s$^{-1}$} and ($6.0 \pm 0.6$) $\times 10^{-14}$ ergs
{cm$^{-2}$} when $N{_H}$ was frozen to $2.8\times 10{^{21}}$ cm$^{-2}$
during the fit.}.

Several quiescent neutron star X-ray transients display a soft thermal
component plus a hard power-law tail in their spectrum (see
e.g. Campana et al. 1998).  Our data do not require a
two-component model. In order to obtain upper limits on the
contribution to the total flux from a possible thermal component, we
fitted the data with this model, consisting of a blackbody or NSA
component and the power-law component, both of them modified by the
effects of interstellar absorption. We fixed the best fit power law
and used 0.2 keV ($10^6$~K) for the blackbody (NSA) temperature as in
Jonker et al. (2005). In this way we find a 95 per cent upper limit to
the fraction of the 0.5-10 keV flux due to a thermal component of $<
15 \%$ ($< 19 \%$).

\section{DISCUSSION}

\subsection{The Outburst Light Curve: Comparison with other Sources}

Fig. 2 shows that the X-ray light curve decays exponentially with a
break or  ``knee'' where the e-folding time becomes faster. A similar
break  has clearly been observed in the decay X-ray light curves of
three other millisecond pulsars: SAX J1808.4-3658 ($P_\mathrm{orb}$=2
hr; Wijnands \& van der Klis 1998, Gilfanov et al. 1998, Wijnands
2005a and references therein), XTE J1751-305 ($P_\mathrm{orb}$=42 min,
Markwardt et al. 2002, Gierli\'nski \& Poutanen 2005) and XTE
J0929-314 ($P_\mathrm{orb}$=43.6 min; Giles et al. 2005, Powell,
Haswell \& Falanga 2006). In the three systems the break  was from an
exponential decay to a linear decay (see Powell et al. 2006). A break
from a slow to a fast exponential decay has been observed in the light
curves of the neutron star X-ray transients Aquila X-1
($P_\mathrm{orb}$=18.95 hr; Maitra \& Bailyn 2004) and Centaurus X-4
($P_\mathrm{orb}$=15.1 hr; Evans, Belian \& Conner 1970, Kaluzienski
et al. 1980; Chen, Shrader \& Livio 1997, Shahbaz et al. 1998). The
initial exponential decay in the light curves of X-ray transients has
been explained by assuming that the evolution of the mass in an
irradiated disk is described as ${\dot{M}_\mathrm{disk}}=-\dot{{M}_c}
\propto {{M}_\mathrm{disk}}$, where $\dot{{M}_c}$ is the central mass
accretion rate. Thus $L{_x} \propto \dot{M}_c =
{{M}^0_\mathrm{disk}} e^{-t/{\tau}_e}$ where ${M}^0_\mathrm{disk}$ is
the initial mass of the irradiated disk. A faster decay phase (steeper
light curve) is expected to occur when the central accretion rate (and
thus the X-ray flux) has decreased below the critical X-ray luminosity
necessary to keep the outer parts of the disk ionized. Then a cooling
front (previously inhibited by X-ray irradiation) will move inwards
the disk and switch off the outburst. The rate of decay in the light
curve during the propagation of the cooling front has been observed to
be either linear or exponential, sometimes showing departures due to
one or more secondary maxima that appear during the outburst
decline. Details on models for the understanding of the outburst light
curves and the outbursts themselves can be found in King \& Ritter
(1998); Dubus, Hameury \& Lasota (2001); Lasota (2001) and Powell et
al. (2006).

The ratio of the optical exponential decay time to the X-ray
exponential decay time ($\tau_e (\mathrm{opt})/\tau_e (\mathrm{X})$) 
is expected to be $\sim 2$ for X-ray transients
when the optical flux is dominated by X-ray reprocessing on the disk
(King \& Ritter 1998). Observational evidence supporting this
prediction is that $< \tau_e (\mathrm{opt})/\tau_e (\mathrm{X}) > \simeq
1.9$ in X-ray novae (Chen et al. 1997). This ratio is $\tau_e
(R)/\tau_e (2.5-25~\mathrm{keV})=0.67 \pm 0.05$ before the knee
in the X-ray light curve of J00291.  Apart from J00291, SAX
J1808.4-3658 is the only millisecond pulsar for which well-sampled
optical and X-ray light curves are available (1998 outburst). For this
system we derive\footnote{$\tau_e(V)=7.4$~d and
$\tau_e(I)=7.9$~d. These values were obtained using the photometric
data from Table 1 in Wang et al. (2001). Only data until MJD 50932.7
were used. $\tau_e (1.5-12 \mathrm{keV})=4.89 \pm 0.06$~d (Powell et
al. 2006).}  $\tau_e(V)/\tau_e (1.5-12~\mathrm{keV})=1.5$. The fact
that the decay in the optical light curve of J00291 is faster than
observed in X-ray irradiated systems suggests that X-ray heating is
insufficient to be the dominant source of optical emission and that
viscous dissipation in the disk may make a significant contribution to
the optical flux. For instance, the optical light curves of dwarf
novae in outburst follow the soft X-rays after the outburst peak
($\tau_e (\mathrm{opt})/\tau_e (\mathrm{X-ray}) \sim 1$) to decline
slower than the X-rays a few days later (Jones \& Watson 1992,
Mauche, Mattei \& Bateson 2001, Wheatley, Mauche \& Mattei 2003).

The knee in the optical light curve of J00291 is delayed $\simeq
4.7$~d with respect to the knee in the X-ray light curve
(Fig. 2). A delay of $\simeq 7$~d was observed in SAX J1808.4-3658 (Wang
et al. 2001). After the knee, the optical and X-ray light curves
of J00291  decay faster with a similar e-folding time, $\tau_e
(R)/\tau_e (2.5-25~\mathrm{keV})=0.8 \pm 0.2$. This was not the case
during the 1998 outburst of SAX J1808.4-3658, when the flux at optical
wavelengths reached a plateau that lasted one month (Wang et
al. 2001). The optical decay in J00291 ($2.2 \pm 0.5$~d mag$^{-1}$;
Section 3) is slower than the $0.93 \pm 0.05 $~d mag$^{-1}$ rate predicted by
the relationship for dwarf novae (non-magnetic CVs) between the rate
of the decay and the orbital period (Bailey 1975). This is given by
Smak (1999) as $({dV/dt})^{-1} = (0.38 \pm 0.02) P_{\mathrm{orb}}
(\mathrm{hr})$. In order to use it we have  assumed $dR/dt \approx dV/dt$
during the decline of J00291. It is interesting to note that a few
intermediate polars (CVs where the accretion disk is disrupted by the
magnetic field of the white dwarf as expected in millisecond pulsars)
have shown dwarf-nova like outbursts which decay faster than the
Bailey's relationship. Examples of these are DO Dra
($P_{\mathrm{orb}}$=3.96 hr, $\check{S}$imon 2000) and HT Cam
($P_{\mathrm{orb}}$=1.35 hr, Ishioka 2002). In summary, J00291 seems to show
during the late decline a longer decay time compared to dwarf novae as
has been observed in other X-ray transients. For instance, the
optical flux of the neutron star transient XTE J2123-058
($P_{\mathrm{orb}}$=5.96 hr) decayed with a rate of $5.00 \pm 0.02$~d
mag$^{-1}$ in the $V$-band and $11.8 \pm 0.3$~d mag$^{-1}$ in the
$R$-band at the end of the 2000 outburst (Soria, Wu \& Galloway 1999,
see also Zurita et al. 2000).

\subsection{The nature of the donor star}

The lack of eclipses or dips in the outburst X-ray light curve imply
an inclination $i\lesssim85^{\circ}$. This limit combined with the
mass function  derived from X-ray data ($2.81311 \times 10^{-5}$
M$_\odot$; Galloway et al. 2005) implies a donor star mass M${_2} \geq
0.04$ (0.05) M$_\odot$ for assumed neutron star mass M$_1$ of 1.4
(2.0) M$_\odot$. Assuming that the inclination of J00291 is drawn from an
isotropic distribution of  inclination angles, using the mass
function and applying the requirement that the companion fits within
its Roche lobe lead to the expectation that the donor star is a
$\lesssim 0.16$ M$_\odot$ (95 \% confidence) low-mass star, most
likely a brown dwarf bloated by the pulsar X-ray emission (Galloway et
al. 2005).

We can use the  peak-to-peak separation in the H$\alpha$ emission
profile ($\Delta V{^{pp}}=650 \pm 40$ km s$^{-1}$) to estimate the
inclination of the system by assuming that the peaks in the averaged
profile represent emission from gas orbiting in the outer radius of
the disk with a Keplerian motion. In such a case $\Delta V{^{pp}} = 2
R_{\mathrm{out}}{\Omega_{\mathrm{K}}} \sin i = 2 (G {M_1} /
{R_{\mathrm{out}}})^{1/2} \sin i$ with $R_{\mathrm{out}}$ being the
outer disk radius. We take $R_{\mathrm{out}}$ to be between the
tidal radius $R_{\mathrm{T}}$ (at which the tidal forces of the donor
star cut the disk off) and the radius of the 3:1 Lindblad resonance
radius $R_{3:1}$ (at which orbits in the disk resonate with the
donor-star orbit, driving the disk
elliptical). ${R_{\mathrm{T}}}\approx 0.9R{_L}(1)$ where $R{_L}(1)$ is the
volume radius of the Roche lobe of the compact star, 
$R_{3:1}={3^{-2/3}}{(1+q)^{-1/3}}a$ where $q = M_2/M_1$ and $a$ is the
separation between both stellar components (see e.g. Whitehurst \& King 1991). The orbital parameters
measured for J00291 allow us to estimate $R{_L}(1)$ as a function of
the mass for the stellar components and $a$ (Eggleton 1983). Using
Kepler's third law together with the constraints on the donor and
neutron star mass we obtain $1.04 < a ({R_\odot}) < 1.19$ and $0.56
{R_\odot} = 3.46 \times 10^{10} \mathrm{cm} < {R_{\mathrm{out}}} <
0.71 {R_\odot} = 4.96 \times 10^{10}  \mathrm{cm}$. From $\Delta
V{^{pp}}$ and ${R_{\mathrm{out}}}$ we estimate a system inclination of
$i=\arcsin (1.145\times10{^{-3}}
({R_{\mathrm{out}}({R_\odot})}/{M_1}({M_\odot}))^{1/2} \Delta
V{^{pp}(km/s))} \simeq 22^{\circ}-32^{\circ}$. This range of inclinations
implies a donor star with mass M$_2=0.04-0.11~ (0.09-0.13)$~M$_\odot$
for a M$_1$=1.4 (2.0)~M$_\odot$ neutron star when using the mass
function of the pulsar derived from X-ray data.

Similarly, an upper limit to the inner radius of the
H$\alpha$-emitting regions during outburst (${R_{\mathrm{in}}}$) can
be obtained by assuming Keplerian motion for the gas. From the maximum
velocity extent of the H$\alpha$ line profile ($FWZI= 2400 \pm 100$ km
s$^{-1}$) we derive ${R_{\mathrm{in}}} \lesssim 4 G {M_1}  (\sin i /
FWZI)^2 = 9.22\times10{^{9}} {M_1}({M_\odot})~{{\sin}^{2}} i$~cm. This is
$\simeq 3000 ~{{\sin}^{2}} i$~times the corotation radius ($R{_{co}}={(G
{M_1} {P{^2}_{spin}}/4{{\pi}^2})}^{1/3}=2.11 \times10{^6}
{M^{1/3}_1}({M_\odot})$~cm, where ${P_{spin}}$ is the neutron star spin
period.

From our photometry during quiescence we find $(R-K) = 4.1 \pm 0.1$
mag, which corrected from extinction corresponds to an unabsorbed
$(R-K)_0$ color of $2.3 \pm 0.3$ mag. We plot in Fig. 6 the
theoretical mass-$(R-K)$ color tracks for low-mass stars with solar
metallicity and ages of 0.1 to 10 Gyr (Baraffe et al. 1998).  The
Gigayear tracks intersect the mass vs $(R-K)$ diagram at $ 0.65
{M_\odot} < M_2 < 0.78 {M_\odot}$.  For these masses the inclination
derived using the mass function is $i\simeq 4^\circ-5^\circ$ which is
highly unlikely given the FWHM  and the double-peaked profiles of the
emission lines, and the fact that the donor star will be much larger
than its Roche lobe (Fig. 6). This discrepancy can be explained if the
donor star and accretion disk are irradiated by a relativistic
particle wind from the pulsar which resumes activity during quiescence
(see e.g. Campana et al. 2004) or by residual accretion onto the
neutron star suface and by thermal X-ray emission from the
pulsar surface which is heated during outbursts (see e.g. Bildsten \&
Chakrabarty 2001).

\subsection{The distance towards J00291}

Different methods have been used to constrain the
distance. First, Galloway et al. (2005) estimated a lower limit of
$\simeq 4$~kpc by assuming  that the 2004 outburst fluence is typical for
the system and that the mass transfer rate is driven by gravitational
radiation.  These authors suggested that the distance cannot be much
larger based on the fact that thermonuclear bursts were not detected
during the outburst event. Secondly, Jonker et al. (2005) derived a
distance of $2.6-3.6$~kpc by assuming that the quiescent X-ray
luminosity of J00291 is similar to that measured for the millisecond
pulsars SAX J1808.4-3658 and XTE J0929-314 in quiescence. Using SAX
J1808.4-3658 alone (for which $d=3.4-3.6$~kpc, Galloway \& Cumming
2006) and the unabsorbed flux of J00291 during quiescence measured in
Section 8, we roughly estimate the distance towards J00291 to be
$2.0-3.4$~kpc. 

The distance to J00291 can also be constrained using the critical X-ray
luminosity necessary in a neutron-star X-ray transient to heat the
disk and produce an exponential decay light curve. We can estimate the
distance to J00291 by assuming that the X-ray irradiation was until
the knee high enough to ionize the whole disk. Shahbaz et
al. (1998) following King \& Ritter (1998) derived expressions for the
critical X-ray luminosity necessary to keep the disk ionized
everywhere. For a neutron star system they found $L{_{\mathrm
crit}}=3.7 \times {10^{36}} R^2_{\mathrm{11}}$ ergs {s$^{-1}$} where
$R_{\mathrm{11}}$ is the ionized disk radius in units of $10^{11}$
cm. The disk radius will be between $R_{\mathrm{out}}$ and the
circularization radius as most of the disk mass will be accreted
during the outburst event. The circularization radius is given by
$R_{\rm circ}/a = (1+q)(R{_L{_1}}/a)^4$ (see e.g. Frank et al. 1992)
where $R{_L{_1}}$ is the distance between the center of the compact
object and the inner Lagrangian point for the donor star. $R{_L{_1}}$
can be approximated in function of $q$ (see e.g. Warner et
al. 1995). Using our constraints on $a$ and $q$ we obtain $2.30 \times
10^{10} \mathrm{cm} < {R_{\mathrm{circ}}} < 3.68 \times 10^{10}
\mathrm{cm}$. From ${R_{\mathrm{out}}}$ and $R_{\rm circ}$ we estimate
$L{_{\mathrm crit}}= 2.0-9.1 \times 10^{35}$ ergs {s$^{-1}$} and from
our fit to the X-ray light curve we derive an unabsorbed X-ray flux of
$\simeq 5 \times 10^{-10}$ ergs {cm$^{-2}$} {s$^{-1}$} at the time of
the knee. Combining both results we find a distance of 1.8 to
3.8 kpc, in line with previous estimates by Jonker et al. (2005).
Taking the apparent $R$- and $K$-band magnitudes during quiescence
(Section 3) and the reddening at these band passes (Section 8), we
derive from the distance module $\log d(\mathrm{pc}) > (5.2 \pm 0.3) -
M_R / 5$ and $\log d(\mathrm{pc}) > (4.7 \pm 0.1) - M_K / 5$. These
are lower limits to the distance towards J00291 as both donor star and
disk are irradiated by the pulsar. In Fig. 7 we show the predicted
mass-absolute magnitude diagram for low-mass stars with ages of 0.1 to
10 Gyr and the mass-distance diagram derived from the distance module.

\subsection{Constraints on the neutron star magnetic field}

Following Burderi et al. (2002) and Di Salvo \& Burderi (2003), we can
place constraints on the neutron star magnetic momentum and thereby the
magnetic field strength $B$ by comparing the X-ray luminosity measured
in quiescence with the expected X-ray luminosity due to residual
accretion onto the neutron star or magnetic dipole radiation. The
X-ray luminosity originating in these processes
depends on both the pulsar spin frequency and $B$ (see Burderi et
al. 2002). Thermal emission from the neutron star may also contribute
to the quiescent X-ray emission (section 9) and therefore these
constraints represent upper limits only. The spin period of the pulsar
in J00291 is 1.67 ms (Galloway et al. 2005). The 0.5-10 KeV quiescent
X-ray luminosity is ${L_x} \simeq 3.4 \times 10^{31} - 1.3 \times
10^{32}$ ergs {s$^{-1}$} (range due to our uncertainty in the
distance). Considering the above processes and using the equations
derived in Di Salvo \& Burderi (2003), the neutron star magnetic field
is most likely less than $3 \times 10^8$~Gauss (a neutron star
with mass 1.4~M$_\odot$ and a radius 10 km was adopted). For
comparison, the magnetic field of the neutron stars in the millisecond
pulsars SAX J1808.4-3658 ($P_\mathrm{spin} \simeq 2.5$~ms), XTE
J1751-305 ($P_\mathrm{spin} \simeq 2.3$ ms) and XTE J0929-314
($P_\mathrm{spin} \simeq 5.4$ ms) are constrained to be $(1-5) \times
10^8$~Gauss, $ < 3 \times 10^9$~(d/10 kpc) Gauss and $< (3-7) \times
10^8$~(d/8 kpc) Gauss respectively (Di Salvo \& Burderi 20003, Wijnands
et al. 2005b).}

\section{CONCLUSIONS}

In this paper we have presented multiwavelength observations of the
millisecond pulsar IGR~J00291+5934. The best source position derived
from the optical and near-infrared images is
$\alpha$(J2000)=00$^h$29$^m$03$^s$.05 $\pm$ 0$^s$.01 and
$\delta$(J2000)=+59$^o$34$'$18$''$.93 $\pm$ 0$''$.05. From the
spectral analysis of our broadband photometry we found strong evidence
for excess in the near-infrared bands that may be due to synchrotron
emission.  We find that the most likely range for the reddening
towards J00291 is $0.7\leq E(B-V)\leq0.9$ ($\equiv 4.06 \times10^{21}$
cm$^{-2}$ $ \leq N{_H} \leq $ $5.22 \times10^{21}$ cm$^{-2}$). The
X-ray spectrum of the source is well-fitted with a power-law model
with photon index $2.4^{+0.5}_{-0.4}$. The unabsorbed quiescent 0.5-10
keV flux is ($7.0 \pm 0.9$) $\times 10^{-14}$ ergs {cm$^{-2}$}
{s$^{-1}$} for $ N{_H} = 4.6\times 10{^{21}}$ cm$^{-2}$. At least 81
\% of the flux in the 0.5-10 keV range is due to this model. The
$(R-K)_0$ color index observed during quiescence supports an
irradiated low-mass donor star and accretion disk contribution. We estimate an inclination of $\simeq
22^{\circ}-32^{\circ}$ based on the H$\alpha$ emission line profile
and we constrain the distance towards J00291 from 2 to 4 kpc. The
magnetic field of the neutron star is most likely $< 3 \times
10^8$~Gauss.

In contrast with the longer recurrence times for X-ray novae, several
millisecond pulsars undergo outbursts in an interval of a few years.
This fact opens the opportunity of obtaining a large sample of
multiwavelength outburst light curves for these X-ray sources, making
possible a future statistical analyses of their light curves as it has
been done for the dwarf novae outbursts. The outburst light curves of
millisecond pulsars may clarify the role that X-ray irradiation plays
in the framework of the thermal instability model both for neutron
star and black hole X-ray transients. Finally, the monitoring of an
outburst should span a large spectral range to allow us to understand
the emission mechanisms in these systems.

\acknowledgements

We thank Jeff McClintock and the anonymous referee for useful
comments on the manuscript. MAPT would like to thank Hans-Jakob
Grimm for guidance on the X-ray analysis. DS acknowledges a
Smithsonian Astrophysical Observatory Clay Fellowship. This work was
supported in part by NASA LTSA grant NAG5-10889 and NASA contract
NAS8-39073 to the Chandra X-Ray Center.  JSB is partially supported
through a Sloan Research Fellowship. The  Peters Automated Infrared
Imaging Telescope (PAIRITEL) is operated by the Smithsonian
Astrophysical Observatory (SAO) and was made possible by a grant from
the Harvard University Milton  Fund, the camera loan from the
University of Virginia, and the continued support of the SAO and UC
Berkeley.  Partial support for the PAIRITEL project was also supplied
by a NASA Swift   Cycle 1 \& 2 Guest Investigator grants. UKIRT is
operated by the Joint Astronomy Center, Hilo, Hawaii, on behalf of the
U.K. Particle Physics and Astronomy Research Council. We would like to
thank the UKIRT Service Observing Programme for obtaining the data.

\clearpage
\begin{figure}
\epsscale{0.8}
\plotone{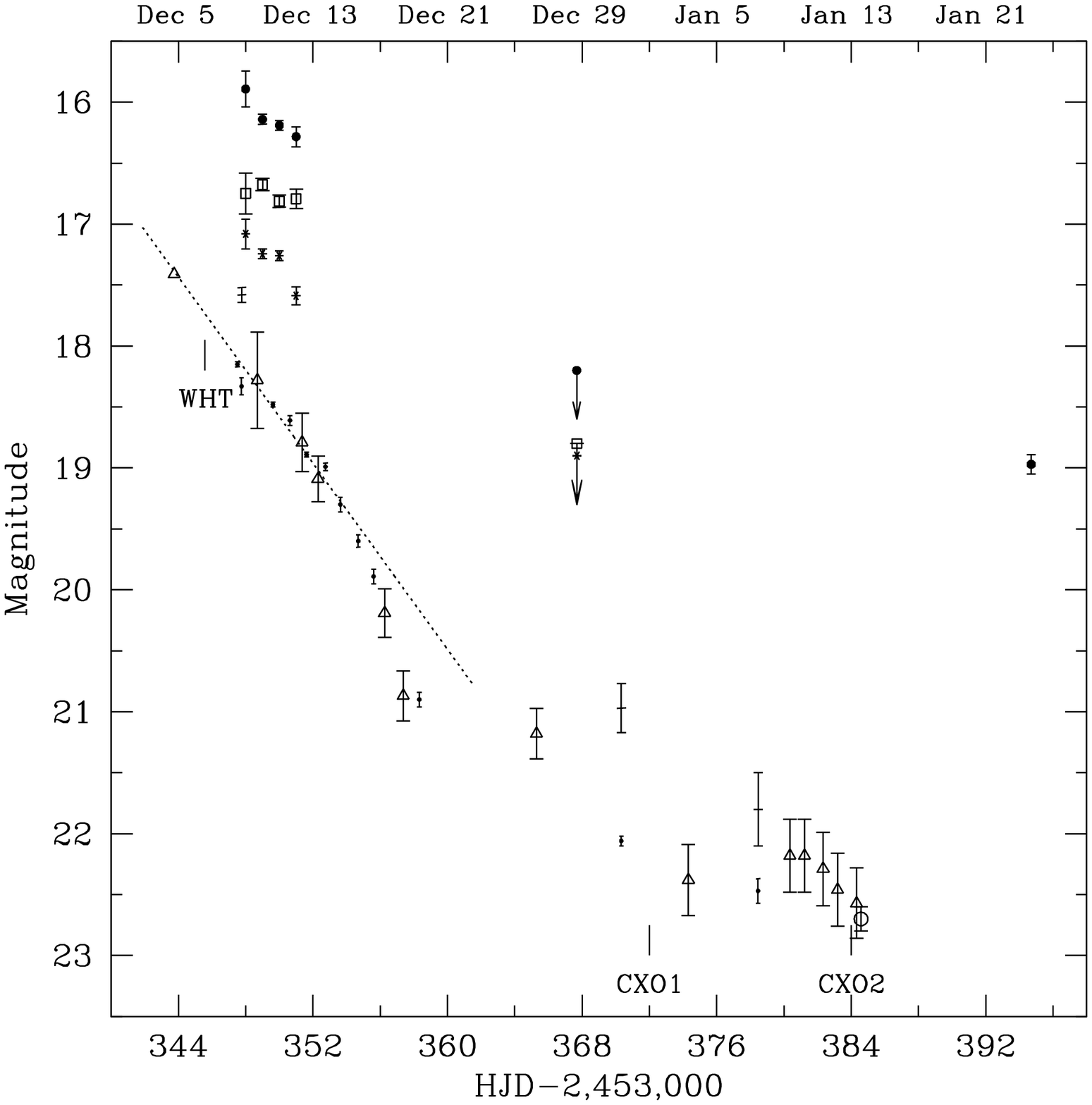}
\caption[]{Optical and near-infrared light curve of J00291 covering the 2004 December-2005 January outburst. Optical and infrared data points are: $R_c$ and $R$ Johnson (dots), $r$ (open circle), $I_c$ and $I$ Johnson (crosses), $J$ ( crosses), $H$ (open squares) and $K_s$ (filled circles). Open triangles mark $R$ band magnitudes reported by Fox \& Kulkarni (2004) and Bikmaev et al. (2005). The latter were taken by digitizing the published light curve. Arrows denote upper limits. The dotted curve shows the first exponential fit described in Section 3. For the sake of comparison this curve is plotted from 2004 December 2 09:00:19 UTC (source discovery with {\it INTEGRAL}) to 2004 December 22 00:00. Times of the WHT spectroscopy during outburst and {\it Chandra X-ray observatory} (CXO) observations during quiescence (Jonker et al. 2005) are marked as WHT and CXO1, CXO2 respectively. The third {\it CXO} observation on MJD 53407.57 is not shown.}
\end{figure}

\clearpage
\begin{figure}
\epsscale{0.8}
\plotone{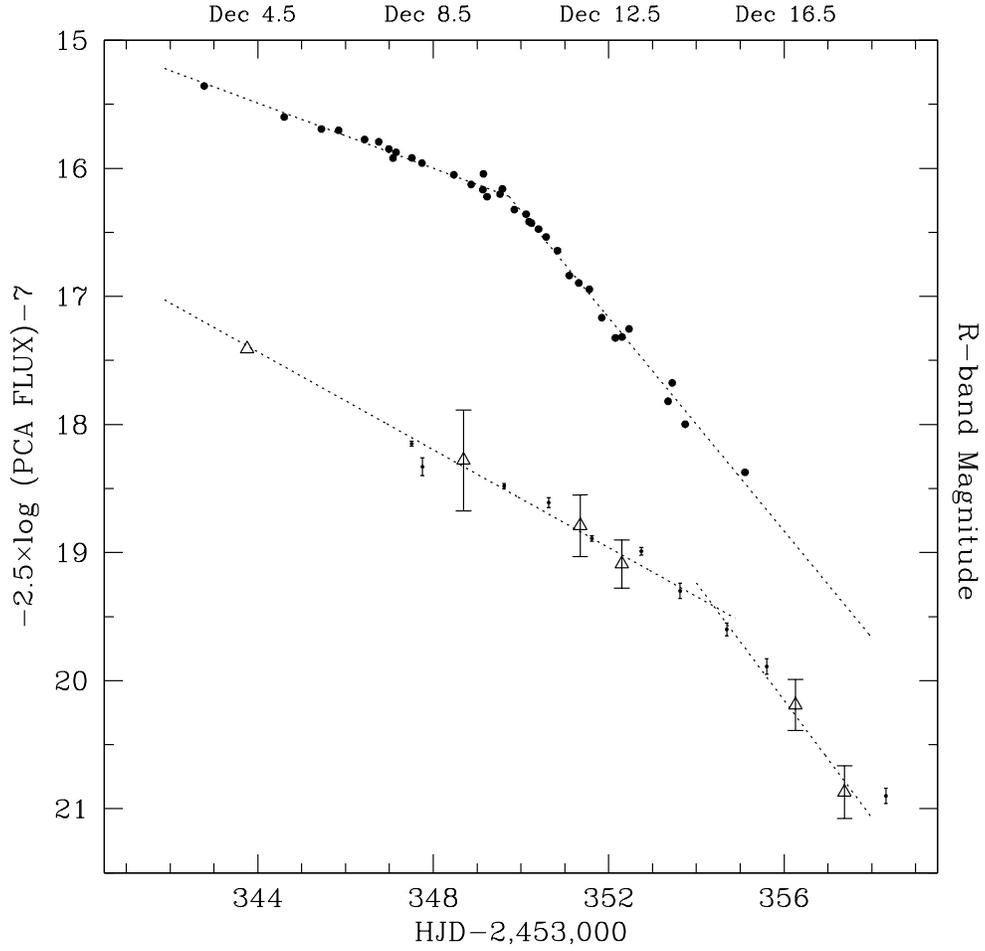}
\caption[]{{\it RXTE/PCA} (2.5-25 keV) light curve (filled circles) and $R$-band outburst light curve (symbols as in Fig. 1) bracketing the 2004 December 2 to 2005 December 19 outburst interval. The dotted curves show the exponential fits to the data. The scale in the Y-axes is the same for the optical and X-ray data. The symbols in the $R$-band (lower) light curve are the same as in Figure 1.}
\end{figure}

\clearpage
\begin{figure}
\epsscale{1.1}
\plottwo{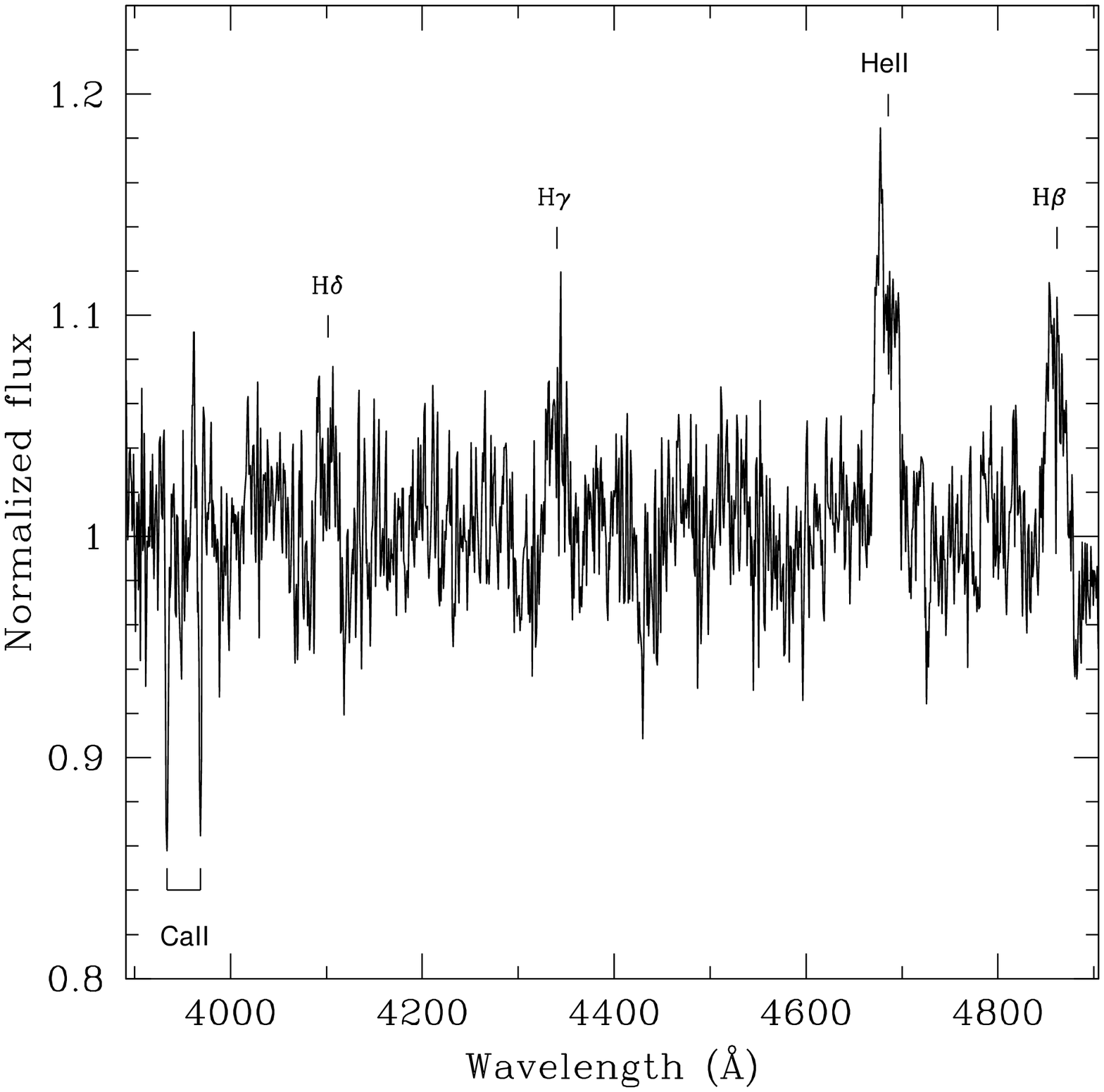}{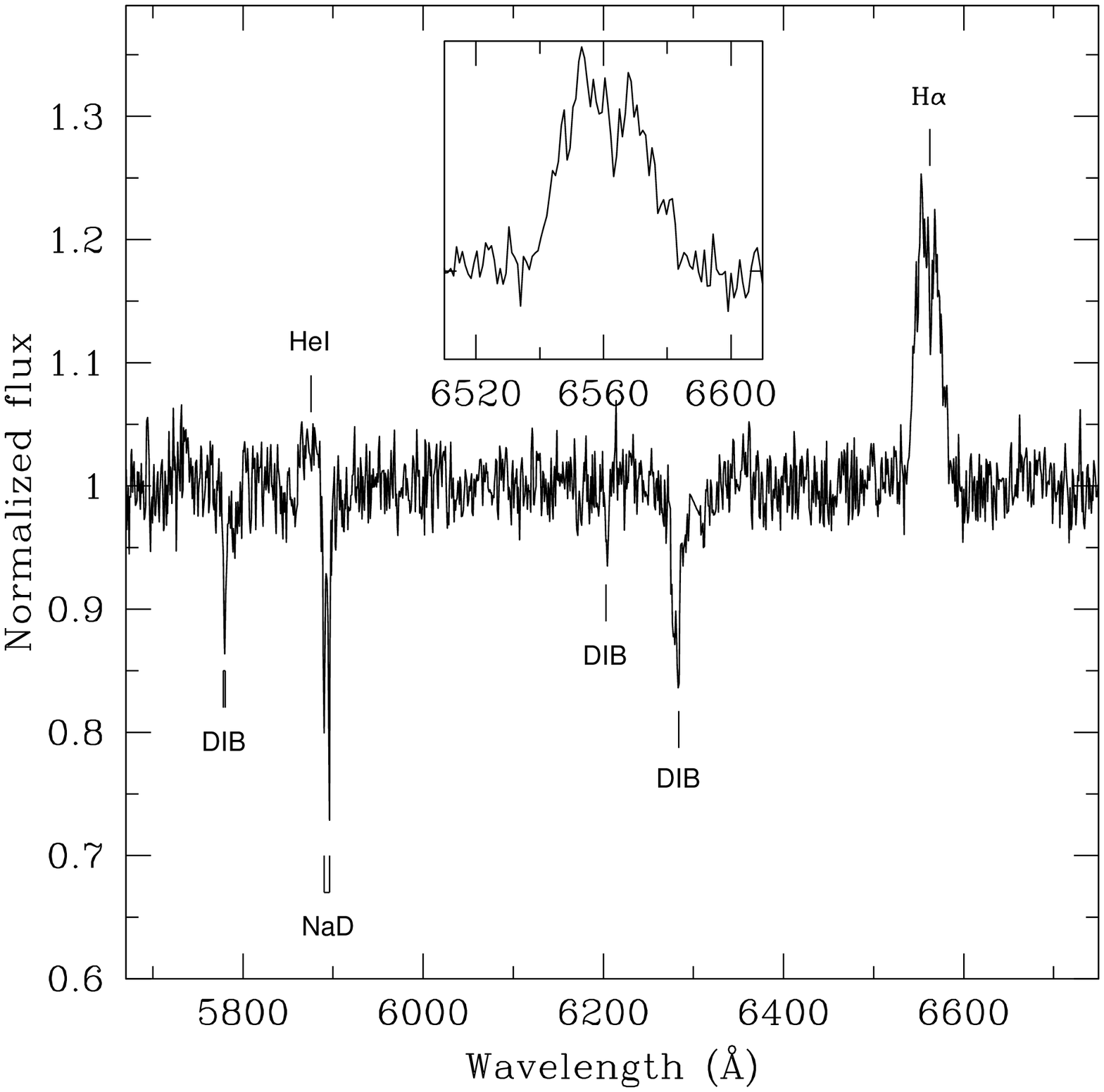}
\caption[]{The normalized and averaged optical spectrum of J00291. Major disk and interstellar features are identified. DIB denotes the diffuse interstellar bands. The inner panel shows a zoom of the averaged H$\alpha$ emission line profile.}
\end{figure}

\clearpage
\begin{figure}
\begin{center}
\includegraphics[width=6.0in]{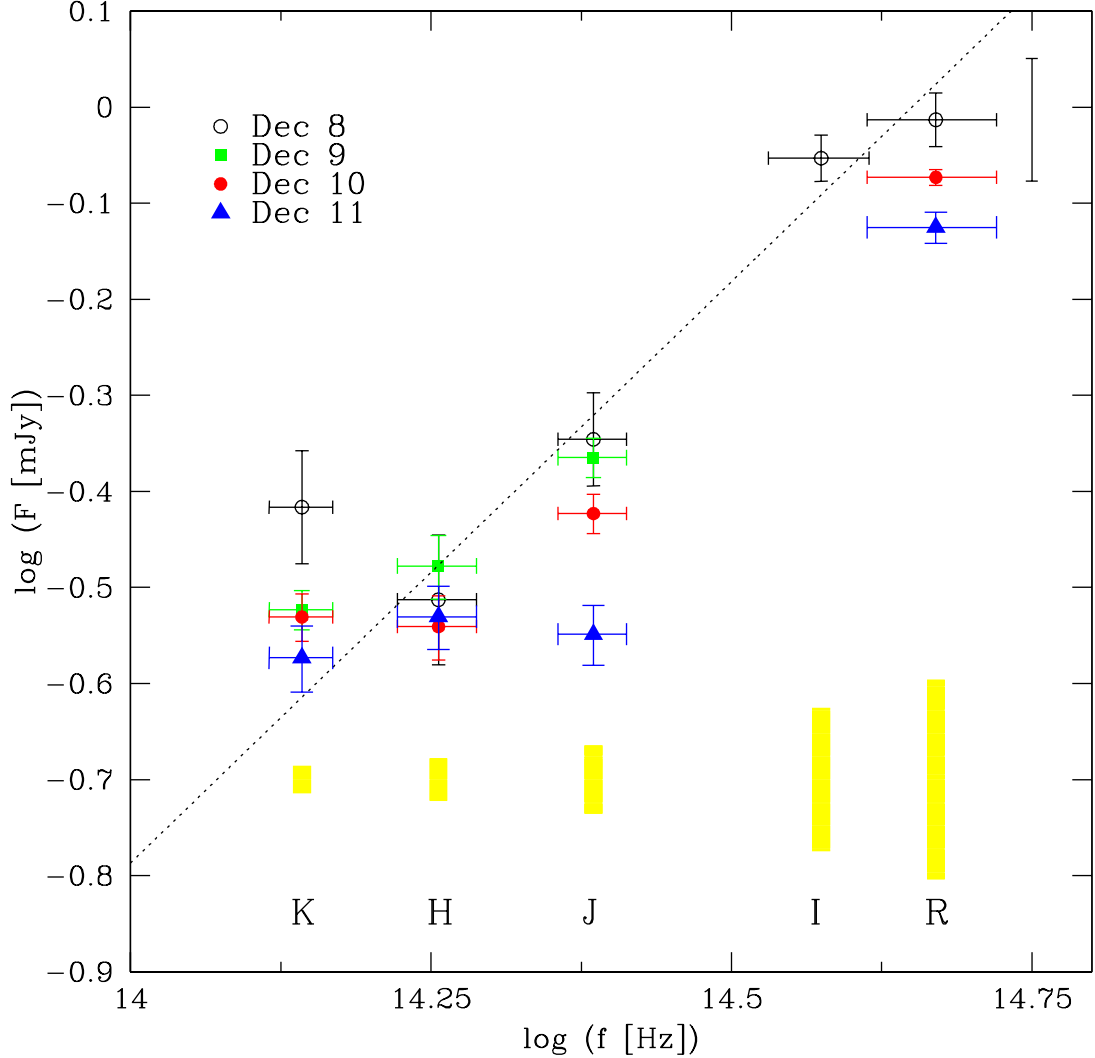}
\caption[]{Optical/near-infrared spectral energy distribution of
J00291 on 2004 December 8, 9, 10 and 11. The fluxes in each band have been
dereddened with E(B-V)=0.8 mag as described in Section 7. The error
bars of the data are the uncertainties in the photometry, whereas the
isolated vertical bar represents the upper limit of the observed flux
variation at optical wavelengths. Horizontal error bars are the $K_s$,
$H$, $J$, $I_c$ and $R_c$ filter bandwidths. Overdrawn as bars
are the effects in the flux (at the central wavelength of each
bandpass) due to the uncertainty ($\pm 0.1$~mag) in the reddening. We
also show the power-law fit to the December 8 data after excluding
the $K$-band flux (dotted line). The fit yields $\alpha = 1.2 \pm 0.2$.}
\end{center}
\end{figure}

\clearpage
\begin{figure}
\begin{center}
\includegraphics[width=4.0in,angle=270]{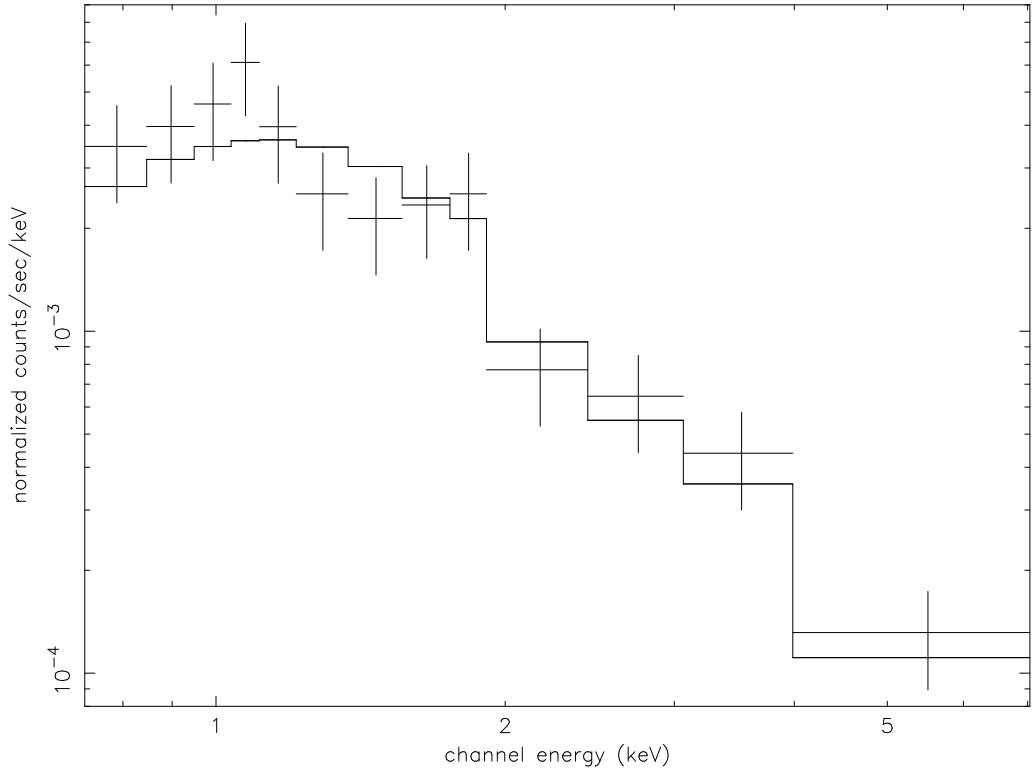}
\caption[]{X-ray spectrum of J00291 obtained during quiescence. The solid line through the spectrum is the best-fit absorbed power-law model with $N{_H}=4.6\times 10{^{21}}$ cm$^{-2}$.}
\end{center}
\end{figure}

\clearpage
\begin{figure}
\epsscale{1.1}
\plottwo{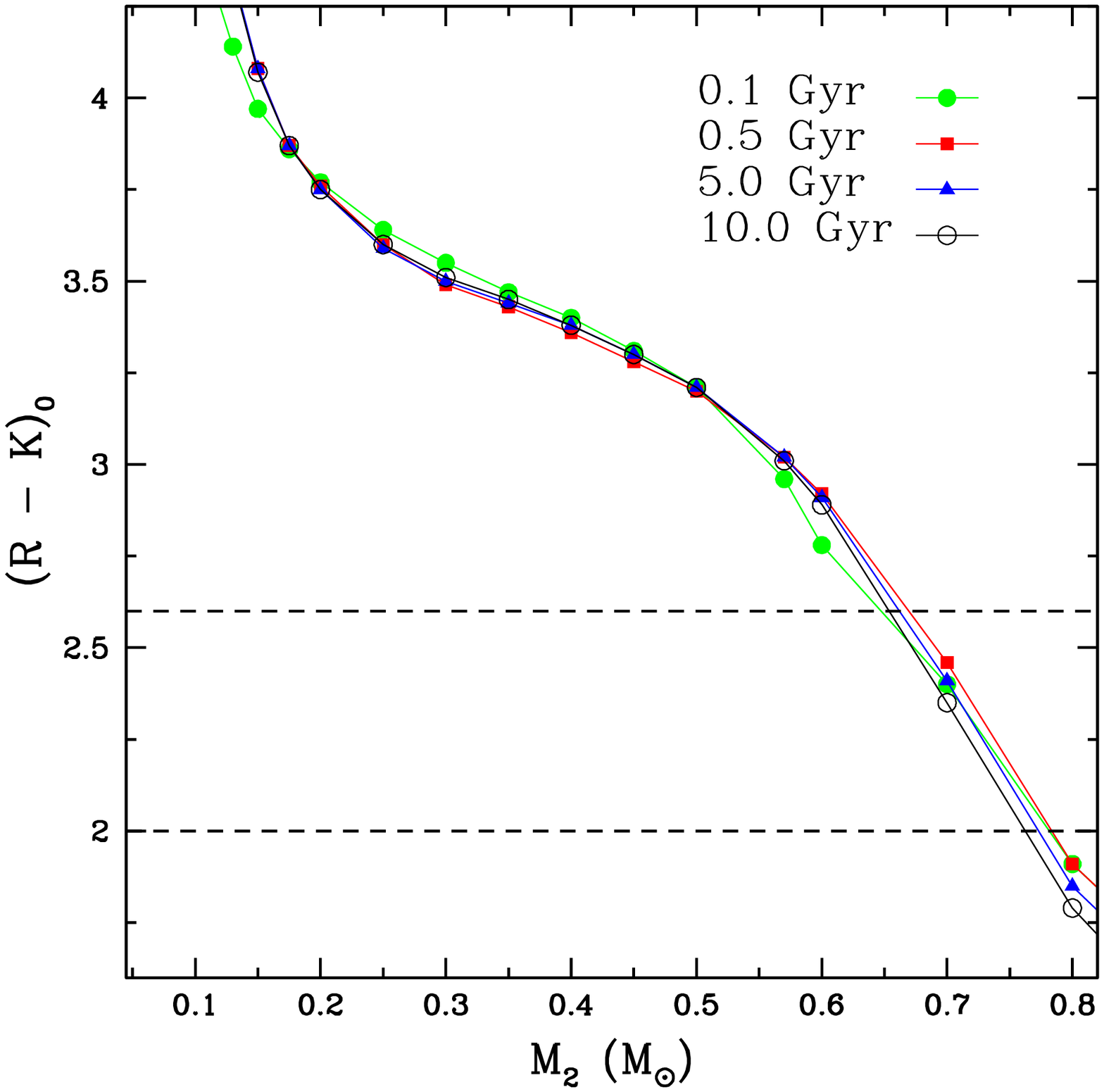}{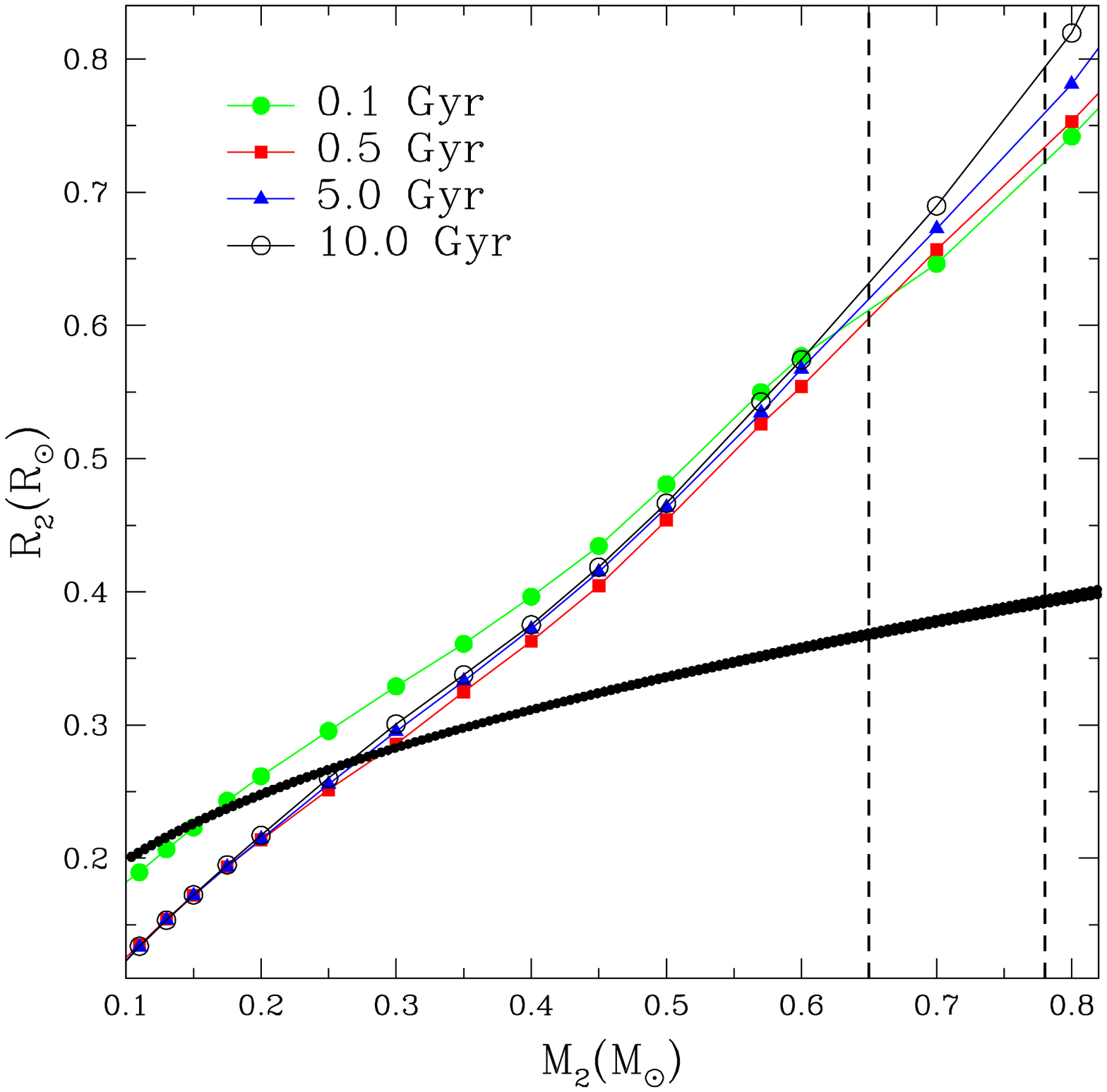}
\caption[]{Left: Predicted color $(R - K)$ as a function of the mass for low-mass stars with ages 0.1, 0.5, 5 and 10 Gyr. Isochrones are taken from Baraffe et al. (1998). The dashed lines delinate our constraint on the $(R - K)$ quiescent color of J00291. Right:  Predicted radius vs mass diagram for low-mass stars as derived from the isochrones presented in the left panel of the figure. The thick solid line represents the Roche lobe of the donor star when adopting a neutron star mass of 1.4 - 2.0~M$_\odot$.}

\end{figure}

\clearpage
\begin{figure}
\epsscale{1.1}
\plottwo{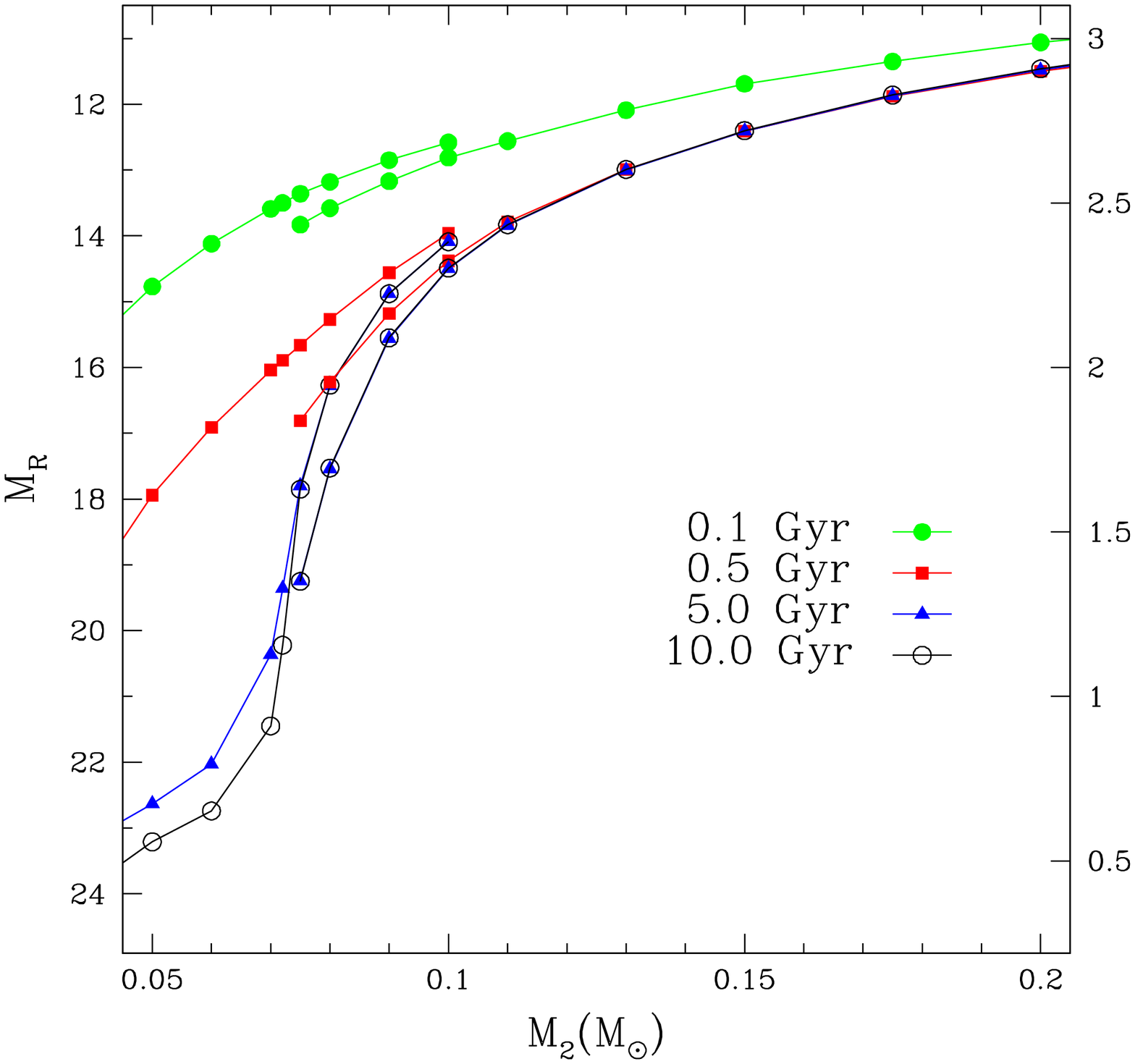}{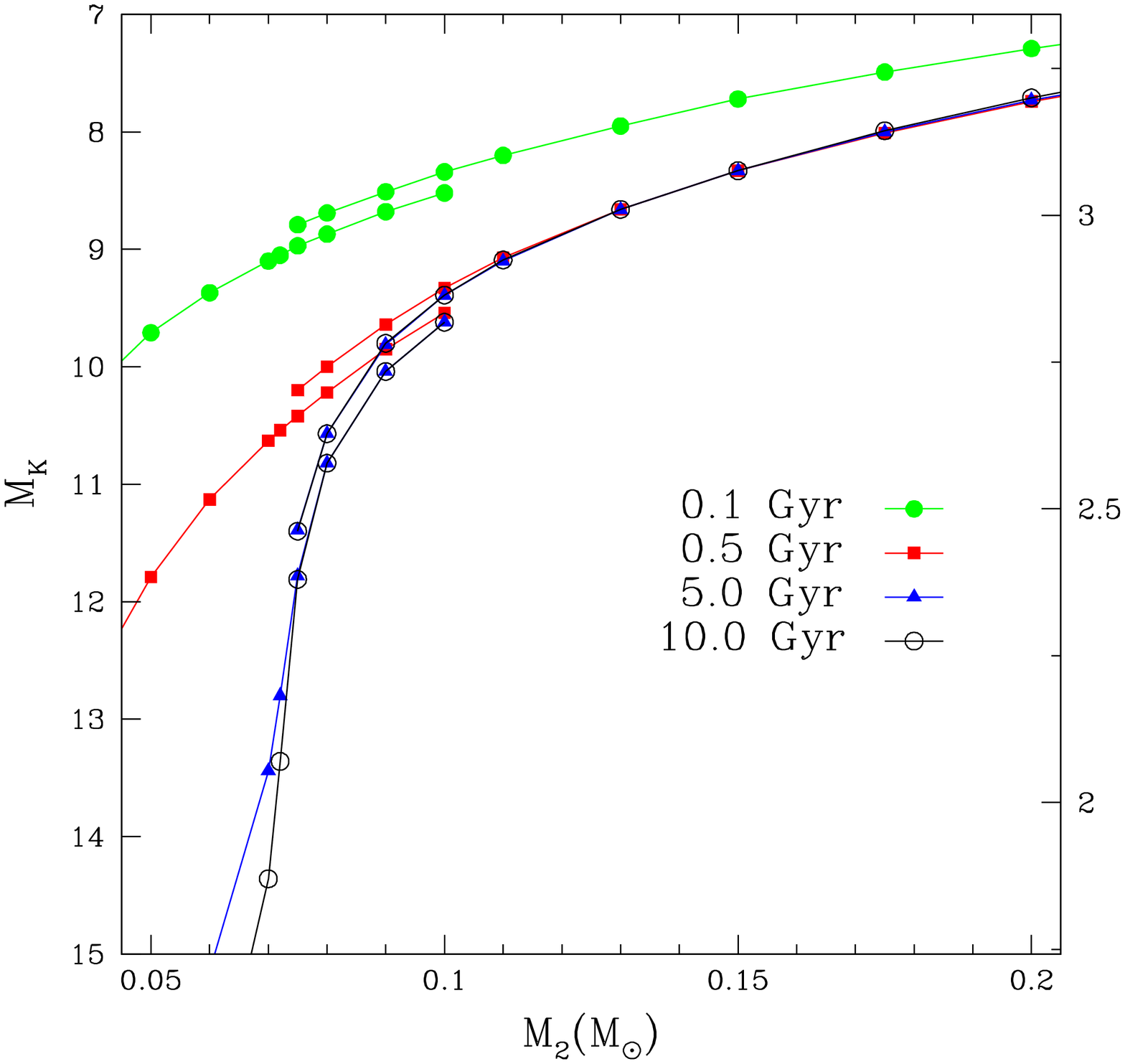}
\caption[]{Predicted $M_R$ (left) and $M_K$ (right) as a function of the mass. Isochrones are taken from Baraffe et al. (1998) for masses $\geq 0.075$~M$_\odot$ and Baraffe et al. (2003) for masses $\leq 0.01$~M$_\odot$. The mismatch between isochrones in the range 0.075~M$_\odot$-0.01~M$_\odot$ reflect model differences. The right-band-side axes are the logarithm of the upper limit to the distance (in pc) towards J00291 given by $5.2 - M_R / 5$ and $4.7 - M_K / 5$ (see Section 10.3).}  

\end{figure}

\clearpage
\begin{deluxetable}{lcccccc}
\tablecaption{Journal of Observations}
\tabletypesize{\scriptsize}
\startdata
\tableline
\tableline
{\em Date } & {\em HJD start--end} & Telescope  & {\em Bandpass/}      &   {\em No.}               & {\em Exp. time} &   {\em seeing}\\
(\em UT)    & (+2,453,000.)        &            & Spectral range (\AA) &   {\em  Images/Spectra}   & {\em (s)}       & (arcsec) \\
\\                            
\\
OUTBURST               &                     &           &       &          &          &         \\
\\
\tableline
\\
\\
04 Dec 5       & 345.39379-345.58101 &WHT        & 3500-5200 & 10 & 830-1500 & $> 1.5$ \\
               &        ,,           &,,         & 5400-7100 & 10 &    ,,    &    ,,  \\
04 Dec 8       & 347.50266-347.51439 & IAC80     & R         & 3  &   300    & 2.1 \\
               & 347.75290-347.76292 & 1.2m      & R,I       & 2,2&   60,60  & 3.2,3.7 \\
               & 347.71597-347.72639 & PAIRITEL  &J,H,K${s}$ & 1  &   439    &     \\
04 ~~~,,~~~9   & 348.56496-348.63947 & PAIRITEL  & ,,        & 3  & 188,424,926  & 2.7 \\
04 ~~~,,~~~10  & 349.60736-349.62958 & 1.2m      & R         & 4  &   300    & 2.4 \\
               & 349.55555-349.69166 & PAIRITEL  &J,H,K${s}$ & 7  & 267-463  & 2.7 \\
04 ~~~,,~~~11  & 350.62887-350.63983 & 1.2m      & R         & 3  &   300    & 2.6 \\
               & 350.55347-350.56666 & PAIRITEL  &J,H,K${s}$ & 1  &   392    &    \\
04 ~~~,,~~~12  & 351.61066-351.62334 & 1.2m      & R         & 3  &   300    & 2.3 \\
04 ~~~,,~~~13  & 352.74011-352.75118 & 1.2m      & ,,        & ,, &      ,,  & 2.0 \\
04 ~~~,,~~~14  & 353.62542-353.63682 & 1.2m      & ,,        & ,, &      ,,  & 3.0 \\
04 ~~~,,~~~15  & 354.68723-354.69858 & 1.2m      & ,,        & ,, &      ,,  & 2.5 \\
04 ~~~,,~~~16  & 355.59923-355.61018 & 1.2m      & ,,        & ,, &      ,,  & 2.8 \\
04 ~~~,,~~~18  & 358.32018-358.32590 & INT       & R         & 3  &   120    & 2.0 \\
04 ~~~,,~~~28  & 367.68095-367.69778 & UKIRT     & J,H,K     & 1  &270,270,540 & 1.4,1.2,1.1\\
04 ~~~,,~~~30  & 370.32417-370.34623 & WHT       & R,I       &7,3 &   180    & 1.2,1.3 \\
05 Jan 7       & 378.45825-378.46899 & TNG       & R,I       &6,2 &   100    & 1.5,1.3 \\
05 ~~~,,~~~14  & 384.57788-384.59612 & MMT       & $r^{'}$   & 3  &   300    & 1.0 \\
\\
\\
QUIESCENCE     &                     &           &      &    &          &         \\
\\
\tableline
\\
05 ~~~,,~~~24  & 394.68693-394.71399        & UKIRT     & K         & 1  &  1620    & 0.7  \\
05 Oct 25      & 669.32713-669.51805 & WHT       & R         &5,20& 300,600  & 1.4,1.2 \\
05 Nov 24      & 698.90954-699.19509 & {\it Chandra}   &0.3-10 keV &1   & 24672    & N/A \\

\enddata
\end{deluxetable}

\clearpage
\begin{table}
\begin{center}
\caption{Emission Line parameters from the average spectrum}
\begin{tabular}{lcccccc}
\tableline\tableline
\\
Emission lines & $\gamma$      & (V$_{b}$+V$_{r}$)/2 & V$_{r}-$V$_{b}$ & FWHM  & FWZI & EW \\    
               & (km s$^{-1}$) & (km s$^{-1}$)       & (km s$^{-1}$)   & (\AA) & (\AA)& (\AA) \\
\tableline 
\\
H$\alpha$       & $-122 \pm 9$  & $-53 \pm 18$  & $650 \pm 40 $ & $30.4 \pm 0.3$ & $52 \pm 3$ & $6.5 \pm 0.4$ \\
H$\beta$        & $-156 \pm 18$ & ---           & ---           &  $22 \pm 1$ & $35 \pm 3$ & $  2.4 \pm 0.3  $ \\
He{\sc ii} $\lambda4686$ & $ -67 \pm 13$$\tablenotemark{a}$     & ---           & ---           & $26 \pm 1$ & $33 \pm 3$ & $2.9 \pm 0.2$ \\  
H$\gamma$       & $ -53 \pm 35$ & ---           & ---           & $24 \pm 2$ & $36 \pm 3$ & $1.7 \pm 0.3$ \\  
H$\delta$       & $-142 \pm 22$ & ---           & ---           & $24 \pm 1$ & $31 \pm 3$ & $1.2 \pm 0.4$ \\   
     
\tableline
\end{tabular}
\tablenotetext{a}{After masking the emission spike on top of the line profile.}
\tablecomments{$\gamma$ designates the positions (shifts) respect to the rest wavelength of the line and was measured with a Gaussian fit as the FWHM. V$_{b}$ and V$_{r}$ designate the shifts respect to the rest wavelength of the line of the blue and red peaks respectively and were measured using the task {\it splot} in {\sc iraf}.}
\end{center}
\end{table}

\begin{table*}[!htb]
\begin{center}
\caption{\label{spetot} Best-fitting parameters of the quiescent spectrum of J00291.}
\begin{tabular}{lcccc}
\hline
      &                   &                              &                       &       \\
Model & Temp./PL index    & Unabsorbed 0.5-10 keV fluxes     & $\chi^2_{\mathrm{red}}$ (d.o.f)& n.h.p \\
      & BB(keV), NSA ($\log K$) & ({10$^{-14}$} ergs {cm$^{-2}$} {s$^{-1}$})   &              & (per cent)      \\
      &                   &                              &                       &       \\

\hline
      &                   &                              &                       &       \\ 
BB    &$0.6 \pm 0.2$      & 3.8  & 3.4 (11) & 0.01 \\
NSA   &$6.5^{+0.4}_{-0.2}$& 3.9  & 3.2 (11) & 0.02 \\
PL    &$2.4^{+0.5}_{-0.4}$& 7.0  & 1.2 (11) & 28 \\
      &                      &       &           &       \\ 
\tableline
\end{tabular}
\tablecomments{NSA, BB and PL
stand for the neutron star atmosphere, blackbody and power-law
models. Quoted uncertainties for the parameters of interest in these
single component models are given at the 90 per cent confidence level
(${\Delta{{\chi}^2}=2.71}$). d.o.f. stands for degrees of freedom and n.h.p. for null hypothesis probability.}
\end{center}
\end{table*}


\begin{thebibliography}{}

\bibitem[]{}Allan, A., Jenness, T., Economou, F., Currie, M. J, \&
Bly, M. 2002, Astronomical Data Analysis Software and Systems XI, ASP
Conference Proceedings, Vol. 281, eds. D. A. Bohlender, D. Durand, \&
T. H. Handley (San Francisco:ASP), 311

\bibitem[]{}Alpar, M. A., Cheng, A. F., Ruderman, M. A., Shaham, J. 1982, Nature, 300, 728 

\bibitem[]{}Bailey, J. 1975, Brit. astr. Ass., 86, 30

\bibitem[]{}Baraffe, I., Chabrier, G., Allard, F., Hauschildt, P. H. 1998, A\&A, 337, 403

\bibitem[]{}Baraffe, I., Chabrier, G., Barman, T. S., Allard, F., Hauschildt, P. H. 2003, A\&A, 402, 701

\bibitem[]{}Beall, J. H., Knight, F. K., Smith, H. A., Wood, K. S., Lebofsky, M., Rieke, G. 1984, ApJ, 284, 745

\bibitem[]{}Bessell, M. S. 1990, PASP, 102, 1181

\bibitem[]{}Bhattacharya, D. \& van den Heuvel, E. P. J. 1991, PhR, 203, 1

\bibitem[]{}Bikmaev, J. et al. 2004, ATel 395

\bibitem[]{}Bildsten, L. \& Chakrabarty D. 2001, ApJ, 557, 292

\bibitem[]{}Blake, C. E. et al. 2005, Nature, 435, 181

\bibitem[]{}Bloom J.~S., Starr D.~L., Blake C.~H., Skrutskie M.~F., Falco E.~E. 2006, in Astronomical Data Analysis Software and Systems XV, ASP Conference Series, Vol. 351, C. Gabriel, C. Arviset, D. Ponz and E. Solano, eds.

\bibitem[]{}Bohlin, R. C., Savage, B. D. \& Drake, J. F. 1978, ApJ, 224, 132   

\bibitem[]{}Burderi et al. 2002, ApJ, 574, 930



\bibitem[]{}Campana, S., Colpi M., Mereghetti, S., Stella, L., Tavani, M. 1998, A\&ARv, 8, 279

\bibitem[]{}Campana et al. 2002, ApJ, 575, 15

\bibitem[]{}Campana et al. 2004, ApJ, 614, 49 

\bibitem[]{}Campana, S., Ferrari, N., Stella, L., Israel, G. L. 2005, A\&A, 434, 9

\bibitem[]{}Cardelli, J. A., Clayton, G. C., Mathis, J. S. 1989, ApJ, 345, 245

\bibitem[]{}Cash W. 1979, ApJ, 228, 939

\bibitem[]{}Cohen, M., Wheaton, WM. A., Megeath, S. T. 2003, AJ, 126, 1090


\bibitem[]{}Chabrier, G., Baraffe, I., Allard, F., Hauschildt, P. 2000, ApJ, 542, 464

\bibitem[]{}Chen, W., Shrader, C. R., Livio, M. 1997, ApJ, 491, 312

\bibitem[]{}Di Salvo, T. \& Burderi, L. 2003, A\&A, 397, 723.

\bibitem[Dickey \& Lockman(1990)]{dic90}Dickey, J. M., \& Lockman, F. J. 1990, ARA\&A, 28, 215

\bibitem[]{}Dubus, G., Hameury, J.-M., Lasota, J.-P. 2001, A\&A, 373, 251

\bibitem[]{}Eckert et al. 2004, ATel 352

\bibitem[]{}Eggleton, P. P. 1983, ApJ, 268, 368

\bibitem[]{}Evans, W. D., Belian, R. D. \& Conner, J. P. 1970, ApJ, 159, L57

\bibitem[]{}Falanga et al.. 2005, A\&A, 444, 15

\bibitem[]{}Fender, R., De Bruyn, G., Pooley, G., Stappers, B. 2004, ATel 361

\bibitem[]{}Fender, R. 2006, in ´Compact Stellar X-ray Sources´, eds. W. H. G. Lewin and M. van der Klis, in press (astro-ph/0303339)

\bibitem[]{}Filippenko, A. V., Foley, R. J. \& Callanan, P. J. 2004. ATel 366 

\bibitem[]{}Fox, D. B. \& Kulkarni, S. R. 2004, ATel 354 

\bibitem[]{}Frank, J., King, A. \& Raine, D. 1992, Accretion Power in Astrophysics (Cambridge: Cambridge Univ. Press) 

\bibitem[]{}Galloway, D. K., Markwardt, C. B., Morgan, E. H., Chakrabarty, D. \& Strohmayer, T. E. 2005, \apj, 622, L45 

\bibitem[]{}Galloway, D. K., Cumming, A. 2006, ApJ, 652, 559

\bibitem[]{}Gierli\'nski, M. \&. Poutanen, J. 2005, MNRAS, 359, 126

\bibitem[]{}Giles, A. B., Greenhill, J. G., Hill, K. M., Sanders, E. 2005, MNRAS, 361, 1180

\bibitem[]{}Gilfanov, M., Revnivtsev, M., Sunyaev, R., Churazov, E. 1998, A\&A, 338, L83

\bibitem[]{}Greenhill, J. G., Giles, A. B. Coutures, C. 2006, MNRAS, 370, 1303 

\bibitem[]{her75}Herbig, G. H. 1975, ApJ, 196, 129

\bibitem[]{her93}Herbig, G. H. 1993, ApJ, 407, 142

\bibitem[]{}Hodkin, S., Irwin, M. \& Hewett, P. 2006, CASU WFCAM/VISTA documentation. No VDF-TRE-IOA-00011-00001

\bibitem[]{}Hynes, R. 2005, ApJ, 623.1026 

\bibitem[]{}Ishioka, R., Kato T., Uemura, M., Billings G. W., Morikawa K., Torii K., Tanabe K., Oksanen A., Hyvönen H., Itoh H. 2002, PASJ, 54, 5811

\bibitem[]{}Jones, M. H. \& Watson, M. G. 1992, MNRAS, 257, 663


\bibitem[]{}Jonker, P. G., Wijnands, R., van der Klis, M. 2004, MNRAS, 349, 94

\bibitem[]{jon05}Jonker, P. G., Campana, S., Steeghs, D., Torres, M. A. P., Galloway, K. K., Markwardt, C. B. \& Chakrabarty, D. 2005, \mnras, 361, 511

\bibitem[]{}Jonker P. G., Galloway D. K., McClintock J. E., Buxton M., Garcia M., Murray S. 2004, MNRAS, 354, 666 

\bibitem[]{}Kaluzienski, L. J., Holt, S. S., Swank, J. H. 1980, ApJ, 241, 779

\bibitem[]{}King A., \& Ritter, H. 1998, MNRAS, 293, 42

\bibitem[]{}Krauss et al. 2005, ApJ, 627, 910

\bibitem[Landolt(1992)]{landolt92} Landolt, A. U. 1992, \aj, 104, 340 

\bibitem[]{}Lasota, J.-P.  2001, NewA Rew, 45, 449

\bibitem[]{}Linares, M., van der Klis, M., Wijnands, R. 2007, ApJ, 660, 595

\bibitem[Lorimer(2005)]{} Lorimer, D. R. 2005, Living Reviews in Relativity, vol. 8, no 7 

\bibitem[]{}Maitra, D. \& Bailyn, C. 2004, ApJ, 608, 444

\bibitem[]{}Markwardt, C. B., Swank J. H., Strohmayer, T. E., Zand, J. J. M int, Marshall, F. E. 2002, ApJ, 575, 21

\bibitem[]{mar04a}Markwardt, C. B., Swank, J. H. \& Strohmayer, T. E. 2004a. ATel 353

\bibitem[]{mar04b}Markwardt, C. B., Galloway, D. K., Chakrabarty, D., Morgan, E. H. \& Strohmayer, T. E. 2004b. ATel 360

\bibitem[]{}Mauche, C. W., Mattei, J. A., Bateson, F. M. 2001, ASP Conf. Ser. 229, in Evolution of Binary and Multiple Stars, ed. P. Podsiadlowski et al. (San Francisco: ASP), 367

\bibitem[]{}McLeod, B. A., Conroy, M., Gauron, T. M., Geary, J. C., \& Ordway, M. P. 2000. Further Developments in Scientific Optical Imaging, 11

\bibitem[]{}Migliari, S., Tomsick, J. A., Maccarone, T. J., Gallo, E., Fender, R. P., Nelemans, G., Russell, D. M. 2006, ApJ, 643, 41 

\bibitem[]{}Nikolaev, S., Weinberg, M. D., Skrutskie, M. F., Cutri, R. M., Wheelock, S. L., Gizis, J. E., Howard, E. M. 2000, AJ, 120, 3340

\bibitem[]{}Nowak et al. 2004, ATel 369

\bibitem[]{pai05}Paizis, A., Nowak, M. A., Wilms, J., Courvoisier, T. J-L., Ebisawa, K., Rodriguez, J. \& Ubertini, P. 2005, A\&A, 444, 357

\bibitem[]{poo04}Pooley, G. 2004, ATel 355

\bibitem[]{}Powell, C. R., Haswell, C. A., Falanga, M. 2006, MNRAS, 374, 466

\bibitem[]{}Radhakrishnan, V. \& Srinivasan, G. 1982, Curr. Sci., 51, 1096 

\bibitem[]{}Ramsay Howatt, S. K et al. 2004, in Proc Spie 5492, UV and Gamma-Ray Space Telescope Systems, eds. Hasinger G, Turner M.J., p.1160


\bibitem[]{}Remillard, R. 2004, ATel 357

\bibitem[]{}Reynolds, M. T. et al. 2006, in Populations of High
Energy Sources in Galaxies. Proceedings of the 230th Symposium of the
IAU, ed. E. J. A. Meurs, G. Fabbiano, Cambrdige: Cambridge University
Press. 230, 80


\bibitem[]{}Roche, P.F. et al. 2003, Proc Spie 4841, Instrument Design and Performance for Optical/IR Ground-Based Telescopes, eds. M Iye and A.F Moorwood)

\bibitem[]{rem04}Roelofs, G., Jonker, P. G., Steeghs, D., Torres, M. \& Nelemans, G. 2004, ATel 356

\bibitem[]{}Rupen, M. P., Dhawan, V., Mioduszewski, A. J. 2004, ATel 364

\bibitem[]{}Russell, D. M., Fender, R. P., Hynes, R. I., Brocksopp, C., Homan, J., Jonker, P. G., Buxton, M. M. 2006, MNRAS, 371, 1334

\bibitem[]{sch05}Schlegel, D. J., Finkbeiner, D. P., Davis, M. 1998, ApJ, 500, 525

\bibitem[]{}Shahbaz, T., Charles, P. A., King, A. R. 1998, MNRAS, 301, 382

\bibitem[]{sha05}Shaw, S. E., Mowlavi, N., Rodriguez, J. et al. 2005, A\&A, 432, L13 

\bibitem[]{}Simons, D. A. \& Tokunaga, A. 2002, PASP, 114, 169 

\bibitem[]{}$\check{S}$imon, V. 2000, A\&A, 360, 627



\bibitem[]{}Skrutskie, M. F. et al. 2006, AJ, 131, 1163

\bibitem[]{}Smak, J. 1999, AcA, 49, 391 

\bibitem[]{}Soria, R., Wu, K., Galloway, D. K. 1999, MNRAS, 309, 528

\bibitem[]{}Steeghs, D., Blake, C., Bloom, J. S., Torres, M. A. P., Jonker, P. G., Starr, D. 2004, ATel 363 

\bibitem[]{}Stetson, P. B.  1987, PASP, 99, 191 

\bibitem[]{}Tokunaga, A. T., Simos, D. A. \& Vacca, W. D. 2002, PASP, 114, 180

\bibitem[]{}Torres et al. 2002, ApJ, 569, 423

\bibitem[]{}Torres et al. 2004, ApJ, 612, 1026


\bibitem[]{}Vrtilek, S. D., Raymond, J. C., Garcia, M. R., Verbunt, F., Hasinger, G., Kurster, M. 1990, A\&A, 235, 162

\bibitem[]{}Wang et al. 2001, ApJ, 563, L61

\bibitem[]{}Warner, B. 1995 in Cataclysmic Variable Stars, (Cambridge University Press: Cambridge)

\bibitem[]{}Wheatley, P. J., Mauche, C. W., Mattei, J. A. 2003, MNRAS, 345, 49

\bibitem[]{}White, N. E. \& van Paradijs, J. 1996, ApJ, 473, 25

\bibitem[]{}Whitehurst, R. \& King, A. 1991, MNRAS, 249, 25

\bibitem[]{}Wijnands, R. \& van der Klis, M. 1998, ApJ, 507, 63

\bibitem[]{}Wijnands, R., Homan, J., Heinke, C. O., Miller, J. M., Lewin, W. H. G. 2005b, ApJ, 619, 492 

\bibitem[]{}Wijnands, R., Homan, J., Miller, J. M., Lewin, W. H. G. 2005, ApJ, 606, 61 

\bibitem[]{}Wijnands, R. 2005a, in Pulsars New Research (NY: Nova Science Publishers), in press, astro-ph/0501264


\bibitem[]{}Zurita, C., Casares, J., Shahbaz, T., Charles, P. A., Hynes, R. I., Shugarov, S., Goransky, V., Pavlenko, E. P., Kuznetsova, Y. 2000, MNRAS, 316, 137
	
\bibitem[]{}Zurita, C. et al. 2002, MNRAS, 334, 999 

\end{thebibliography}
\end{document}